\documentclass[%
 reprint,
superscriptaddress,
 amsmath,amssymb,
 aps,
 twocolumn,
]{revtex4}

\usepackage{graphicx}
\usepackage{dcolumn}
\usepackage{bm}
\usepackage{color}

\begin{document}
\title{Solid-Liquid Composites for Soft Multifunctional Materials}

\author{Robert W. Style} 
\email{robert.style@mat.ethz.ch}
\affiliation{Department of Materials, ETH Z\"urich, 8093 Z\"urich, Switzerland.}

\author{Jin Young Kim} 
\affiliation{Department of Materials, ETH Z\"urich, 8093 Z\"urich, Switzerland.}

\author{Ravi Tutika} 
\affiliation{Soft Materials and Structures Lab, Mechanical Engineering, Virginia Tech, Virginia, USA.}

\author{Michael D. Bartlett} 
\email{mbartlett@vt.edu}
\affiliation{Soft Materials and Structures Lab, Mechanical Engineering, Virginia Tech, Virginia, USA.}

\affiliation{Department of Materials, ETH Z\"urich, 8093 Z\"urich, Switzerland.}

\begin{abstract}
Soft materials with a liquid component are an emerging paradigm in materials design. The incorporation of a liquid phase, such as water, liquid metals, or complex fluids, into solid materials imparts unique properties and characteristics that emerge as a result of the dramatically different properties of the liquid and solid.
Especially in recent years, this has led to the development and study of a range of novel materials with new functional responses, with applications in topics including soft electronics, soft robotics, 3D printing, wet granular systems and even in cell biology.
Here we provide a review of solid-liquid composites, broadly defined as a material system with at least one, phase-separated liquid component, and discuss their morphology and fabrication approaches, their emergent mechanical properties and functional response, and the broad range of their applications.
\end{abstract}

\newcommand\edits[1]{\textcolor{black}{#1}}
\newcommand\sifig[1]{\textcolor{black}{#1}}

\maketitle

\section{Introduction}

\begin{figure*}[th!]
 \centering
 \vspace{-5pt}
	\includegraphics[width=1\textwidth]{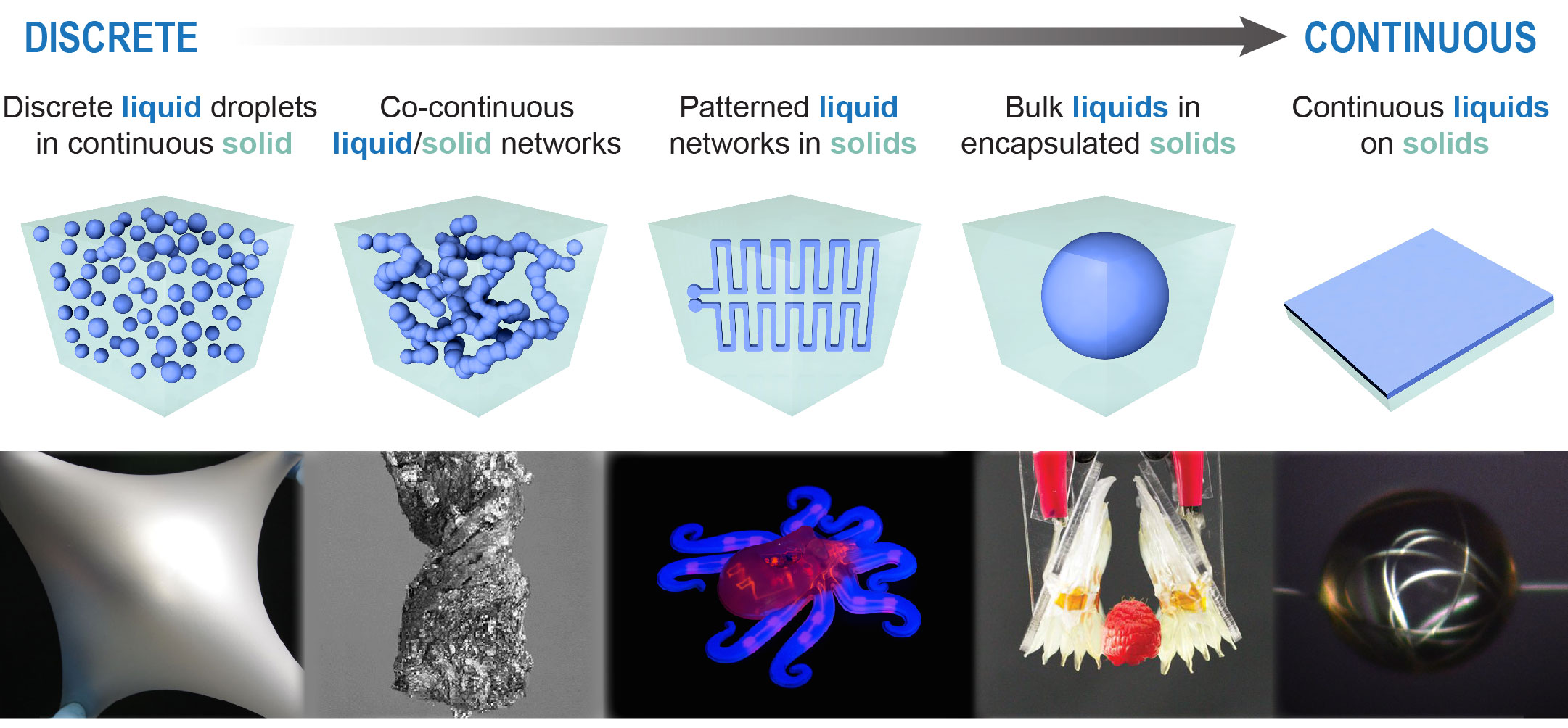}
    \vspace{-5pt}
	\caption{The range of different types of solid-liquid composites. 
	{\bf{Small, discrete liquid inclusions in a continuous solid}} (e.g. LM inclusions in soft elastomers).\cite{Bartlett2016} \copyright (2016), National Academy of Sciences.; 
	{\bf{Co-continuous liquid/solid networks}} (e.g. morphing metal and elastomer bicontinuous foams). Reproduced with permission\cite{van2016morphing} \copyright2016, Wiley-VCH; 
	{\bf{Patterned liquid networks in solids}} (e.g. entirely soft, autonomous robot \textit{octobot}). Reproduced with permission.\cite{wehner2016integrated} \copyright 2016, Springer.; 
	{\bf{Bulk liquids in encapsulated solids}} (e.g. self-healing electrostatic actuators). From\cite{acome2018hydraulically}, reprinted with permission from AAAS.; 
	{\bf{Continuous liquids on/in solids}} (e.g. in-drop capillary spooling inspired hybrid fibers with mixed solid–liquid mechanical properties).\cite{elettro2016} \copyright (2016), National Academy of Sciences.
	}
    \label{fig:types}
    \vspace{-5pt}
\end{figure*}

Solid-liquid composites (SLCs) represent an emerging area where liquid components are incorporated into solid materials to enable new properties.\cite{bartlett2020introduction}
Although the interaction between a fluid and a solid has a long history of study (e.g. wetting theory  \cite{RevModPhys.57.827}), recent work has shown how novel combinations of soft materials and fluid can provide unique and enabling behaviour.
For example, this can be achieved by adding functional liquids, such as liquid metals (LMs) and ferrofluids, to soft elastomers.
This addition imbues the composite with the liquid's properties, such as electrical, thermal, magnetic, or actuation response \edits{(which are typically weak in soft solids)}, while maintaining the soft and deformable properties of the elastomer phase.
SLCs can take a wide range of forms, as shown in Figure \ref{fig:types}, ranging from discrete droplets in continuous solids, co-continuous and patterned liquid networks is solids, and bulk liquid inclusions embedded in and on solid materials.
Furthermore, these can comprise a huge range of materials.
For example, when liquid metal is dispersed as droplets into soft solids, electronics and machines can be created that can be stretched, bent, and even cut, ruptured, or have sections removed while remaining functional.\cite{Bartlett2017,Markvicka2018}
Ferrofluids and phase change materials can be embedded in solids to impart dynamic change of rigidity and shape.\cite{van2016morphing,testa2019}
Furthermore, patterned microchannels filled with reactive liquids can enable autonomous actuation of soft robotics,\cite{wehner2016integrated}
liquid-filled reservoirs can be activated electrically to enable dynamic actuation for lifting or grasping,\cite{acome2018hydraulically}
while fluid droplets on thin elastic fibers can enable self-assembled spooling for highly deformable structures.\cite{elettro2016}
In some cases, the combination even brings unexpected novel properties, where liquid droplets can impart extreme toughness through reconfiguration of droplets,\cite{kazem2018} or increases the stiffness of soft solids due to the surface tension between the solid and liquid.\cite{style2015}

The enabling properties of SLCs find a wide range of applications  across soft matter engineering.
Due to their soft yet functional response, they are ideal for soft robotics, soft actuation, and soft electronics where soft components replace traditionally rigid components to achieve high levels of deformability and compliance.\cite{rus2015design, majidi2014soft}
Beyond these topics, there are a whole host of other uses, ranging from switchable adhesives,\cite{croll2019switchable} to drug delivery,\cite{singh2014} creating colourful materials,\cite{style2018liquid} compartmentalising chemical reactions,\cite{faccio2019} and active materials.
A constant source of inspiration is the example of Nature.
Many biological materials derive from liquid composites, and Nature has used these to achieve a range of properties, from beautiful structural colours,\cite{dufresne2009self} to cellular regulation,\cite{shin2017} slippery surfaces,\cite{bohn2004} and materials with excellent mechanical properties.\cite{niklas1992plant}
Note that there are also many examples of hydrogels with interesting properties.
However, we shall not discuss these in detail, but consider only materials containing distinct, \edits{phase-separated}, liquid inclusions, \edits{that are large enough to have the same, uniform properties as they have in bulk (i.e. the inclusions are much bigger than the molecular scale, or the scale of any colloidal components)}.

With this review, we aim to give an introduction to the field in three main sections.
We first highlight the range of techniques that can be used to fabricate SLCs, including details of common material combinations, material morphologies, and important constraints.
Second, we discuss the range of properties that can be achieved with SLCs.
Third, we review how such materials are used for applications ranging from actuation and soft robotics to electronics.
Finally, we discuss current challenges and future opportunities.

\section{Fabrication and Morphology}

\begin{figure*}
 \centering
 \vspace{-5pt}
	\includegraphics[width=1\textwidth]{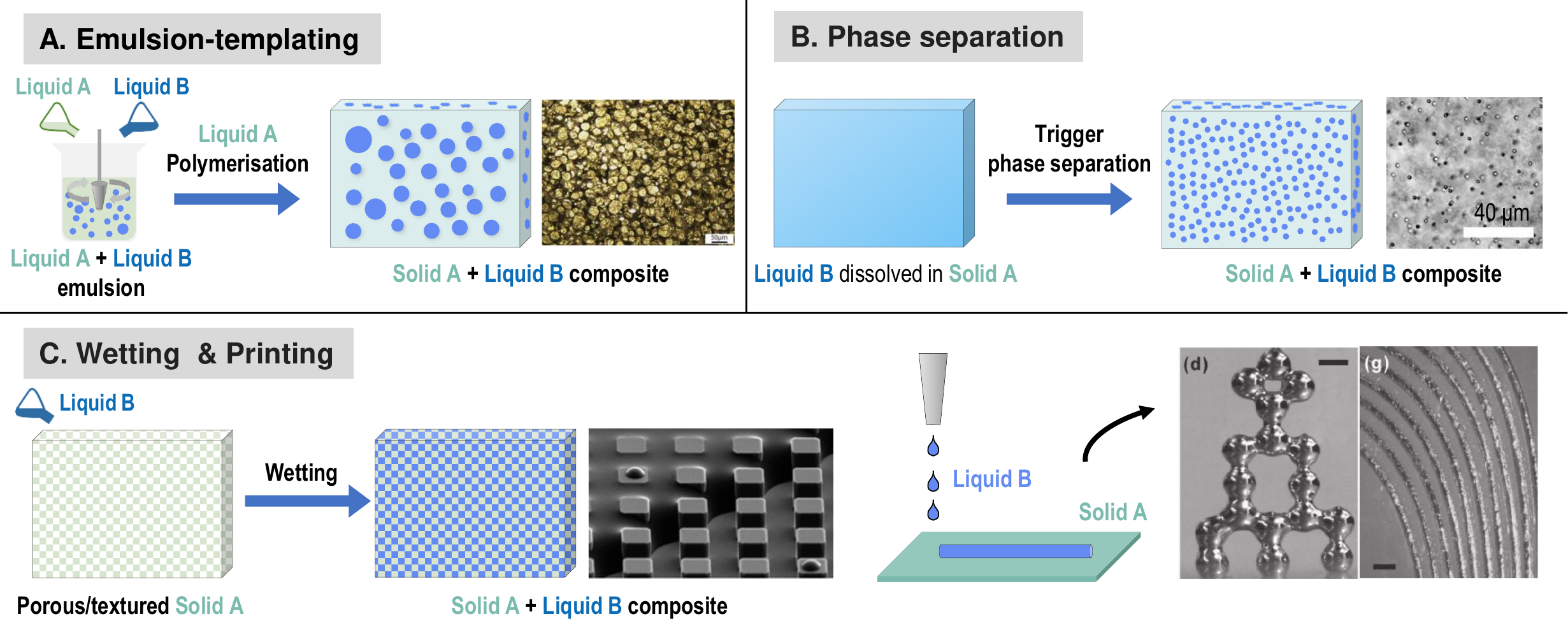}
    \vspace{-5pt}
	\caption{The three overarching types of SLCs. 
	A) Composites made by solidifying one component of an emulsion. The image shows liquid metal droplets in a silicone elastomer, made by emulsion templating \cite{Tutika2019liquid}. Reprinted with permission from \copyright (2019) American Chemical Society.
	B) Composites made via phase separating droplets droplets in a polymeric solid. The image shows anise-oil droplets that have formed in a soft, polydimethylacrylamide hydrogel \cite{style2018liquid}. 
	C) Composites made via wetting and printing on preformed solids. An ionic liquid droplet wets on a textured solid \cite{smith2013droplet} and 3D printing of free standing LM \cite{ladd20133d}. Reproduced with permission from \copyright (2013) The Royal Society of Chemistry and \copyright (2013) Wiley-VCH.}
    \label{fig:fabrication}
    \vspace{-5pt}
\end{figure*}

Although there are a range of different ways of making SLCs, broadly speaking, they can be subdivided into three overarching groups. This is shown schematically in Figure \ref{fig:fabrication} and organized as:
\begin{itemize}
    \item Emulsion-templating: made by solidifying one component of a (typically immiscible) emulsion.
    \item  Phase separation: made when the liquid phase forms via phase separation.
    \item Wetting and printing: adding liquids to preformed solids
\end{itemize}
For each of these, there are a number of important considerations, and here we give a brief overview.
In particular, we cover liquid choice in section \ref{sec:liquidchoice}, and then for each type of fabrication approach (sections \ref{sec:emulsions}-\ref{sec:wetting}), we discuss solid-liquid compatibility constraints, and important aspects of fabrication.
\edits{Our discussion also includes the size scale of the liquid inclusions which can be controlled by the fabrication approach.}
\edits{Note that these three groups are not exhaustive, and there are alternative approaches that are variations on these techniques. For example, the wetting approach can be modified by infiltrating a curing polymer into a porous, low-melting-temperature solid, to create a solid composite, and then melting the solid. \cite{yang2020ultrasoft}}

\subsection{The choice of liquid \label{sec:liquidchoice}}

The liquid component of the composite often imparts the desired functionality. The liquid selection is therefore an important initial decision. 
Fortunately, depending on the required outcome, there are often a range of suitable liquids.
This is illustrated in Figure \ref{fig:miscibility}, which shows, broadly speaking, functionalities that can be achieved with liquids from six broad categories: metallic, water-based, hydrocarbons, silicones, flurocarbons, and ionic liquids. This is certainly not an exhaustive list of options, but covers a large selection of commonly available liquids.

When magnetic response is desired, there are a range of different magnetic liquids available, based on different liquid chemistries.
Practically speaking, these generally consist of iron particles, stably suspended in a carrier liquid to form a ferrofluid or magnetorheological fluid (MRF) -- normally with a surfactant to stabilise the particles against aggregation.
Ferrofluids have smaller particles that do not sediment over time, while MRF's have larger particles, which can sediment, but yield a much larger change in rheological properties upon the application of a magnetic field.
Most commonly, carrier liquids are water or a hydrocarbon, and these are commercially available.
However, silicone oils and fluorinated oils, \cite{wang2018} LMs,\edits{\cite{hu2019,cao2020ferromagnetic}} and ionic liquids \cite{jain2011stable} have all been demonstrated as functioning alternatives.

For electrically-conductive liquids, options include LMs, ionic liquids, and salt solutions.
Of these, LMs have the highest conductivity,\cite{Chiechi2008ACIE} $\sigma= O(10^6\mathrm{S}/\mathrm{m})$, while ionic liquids and salt solutions can have conductivities up to $\sigma=O(1\mathrm{S}/\mathrm{m})$ (e.g. \cite{zech2010}).

For thermally-conductive liquids, LMs offer the highest conductivity, with $k=O(10-50\mathrm{W/mK})$.
Other liquids have significantly lower conductivity, with water having $k=0.6\mathrm{W/mK}$, and ionic liquids and many hydrocarbons being in the range $k=0.1-0.3\mathrm{W/mK}$.
Further, this useful property of the LMs is due to their mobile electrons, which transport thermal energy freely through the material.
MRF's also have increased $k$ -- a few times larger than $k$ for their carrier liquid -- due to their high loading with iron particles.
They have also been seen to exhibit an increase in conductivity upon application of a magnetic field, as particles in the MRF align to form conductive chains. \cite{cha2010}

To create composites \edits{that are tough in response to static loadings}, thus far, only LMs have shown a generic toughening effect on the properties of a soft composite,\cite{kazem2018,peng2019} although recent work has suggested a potential use of nanodroplets of high-boiling point oligomers.\cite{wang2020toughening}
However, a key component of of toughening is introducing dissipative components into a material.
We note that this can be done by introducing viscoelastic, or yield-stress inclusions (such as an MRF when magnetised) into a composite \edits{ and so there may be possibilities to create statically-tough materials via alternative approaches.
Such dissipative liquid inclusions certainly can improve toughness in response to dynamic loadings, for example in terms of their resistance to ballistic penetration. \cite{lee2003ballistic}}'

In many applications, the fluid is also required to be chemically inert.
This is particularly important for biomedical applications, or when the liquid is exposed to the environment on the surface of the composite, as with lubricant-infused surfaces.\cite{wong2011,solomon2016}
In this case, fluorinated oil and silicone oil are well suited.
Both are broadly biocompatible (e.g. \cite{malchiodi2002biocompatibility}), chemically stable, and non-flammable.
Fluorinated oil is particularly useful as it is insoluble in most other liquids -- both polar and non-polar -- including alcohols, alkanes, water, acids and bases.
Thus it won't absorb into foreign materials touching the surface of the composite (e.g. \cite{chen2020design}).
By contrast, silicone oil is insoluble in water and short chain alcohols, and relatively resistant to acid and alkaline solutions, but is highly soluble in solvents like hexane or toluene.\cite{chen2020design}

One overarching requirement is that the liquid needs to be stable. For this it should ideally have a low volatility so that it will not diffuse out through the solid and evaporate into the surrounding environment.
Out of the materials presented in Figure \ref{fig:miscibility}, ionic liquids, LMs and silicone oils typically have low vapour pressures, with ionic liquids in particular having extremely low vapour pressures -- even nanodroplets will not evaporate at observable rates.\cite{heim2013measurement,sedev2011surface}
On the other hand, water, hydrocarbons and fluorinated oils are reasonably volatile.
In this case, care needs to be taken that the liquids need to be completely sealed off inside relatively impermeable solids, or stabilised, for example by adding a nonvolatile solute.
For the case of water, it can be stabilised against evaporation by addition of sufficiently high concentrations of glycerol or a salt.

\subsection{The choice of solid \label{sec:solidchoice}}
\edits{Once the liquid has been chosen, a suitable solid can be selected.
This choice will typically be dominated by the compatibility requirements that we discuss below.
However, further requirements like stiffness, extensibility, fracture toughness, temperature stability, or price can be easily selected using standard Ashby diagrams.\cite{ashby2018materials}
For soft materials the solid generally consists of elastomer or gel, with common material choices including silicones, polyurethanes, thermoplastic elastomers like SIS rubber, and hydrogels.
These materials represent both chemically and physically crosslinked systems.
}

\subsection{Emulsion-templated materials \label{sec:emulsions}}

\begin{figure}
 \centering
 \vspace{-5pt}
	\includegraphics[width=9cm]{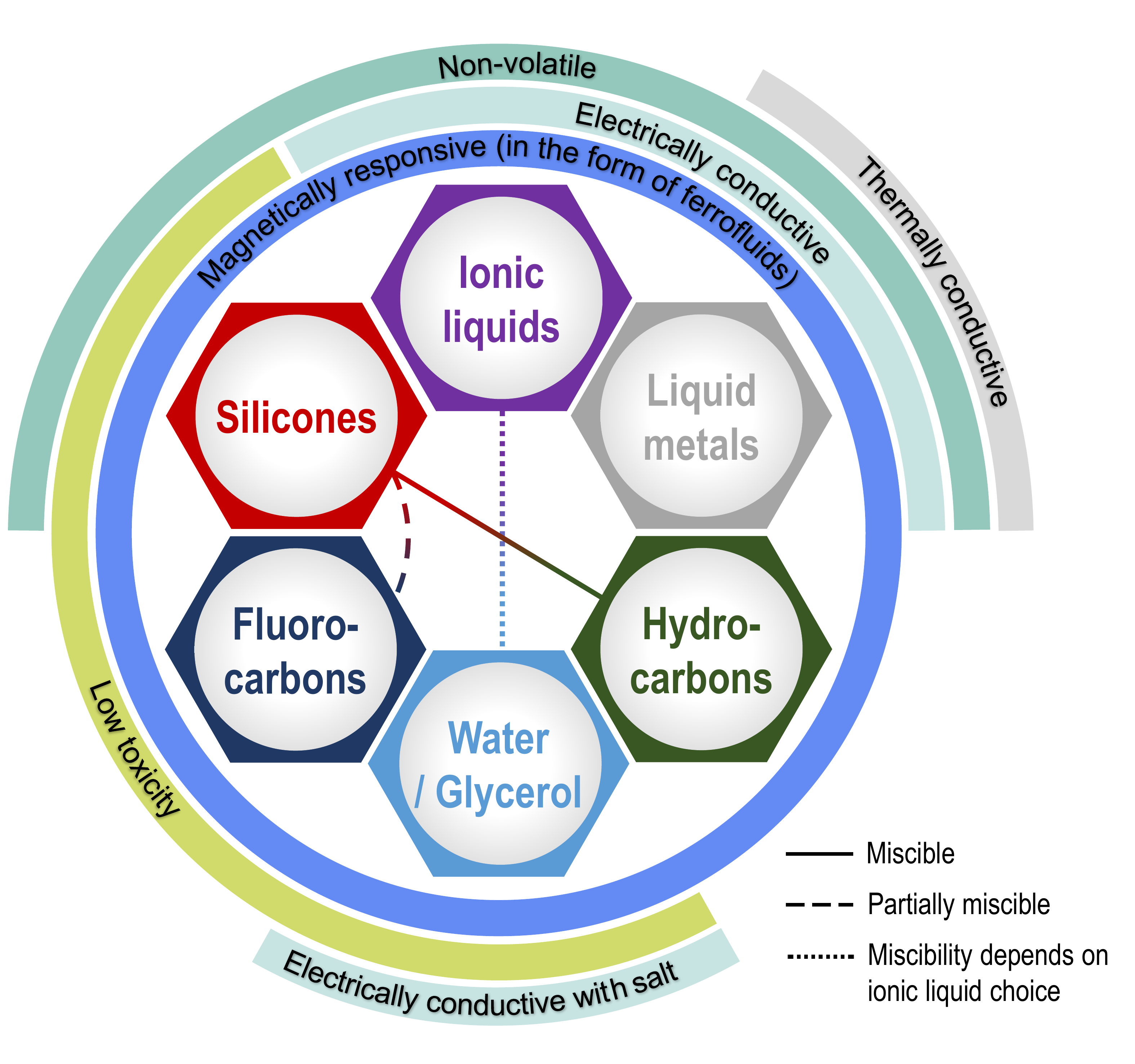}
    \vspace{-5pt}
	\caption{General miscibility/characteristics of a selection of six different common material classes for making SLCs.
		The outer rings indicate which typical characteristics SLCs display.
	No line between classes indicate that they are \emph{generally} immiscible.
	A solid line indicates that two different classes are miscible (silicones$-$hydrocarbons \cite{rochow1951introduction}). A dashed line between them indicates partial miscibility \edits{(e.g. the solubility of the fluorinated oil Fluorinert FC770 in a silicone gel is 3.5$\%$ \cite{rosowski2020elastic})}. Ionic liquids and water/glycerol pairs can vary from completely miscible to immiscible, depending on the choice of ionic liquid.\cite{yokozeki2010water,anand2012}
	Note that most hydrocarbons and fluorocarbons are immiscible, but there have been observations of partial miscibility (e.g. \cite{sett2017lubricant}).
	Note also, that `immiscible' pairings always have a small, and sometimes significant miscibility. For example, the solubility of benezene in $20^\circ$C water is 0.17$\%$, \cite{arnold1958solubility}, while water has a very low solubility in silicone, but can still diffuse through it at a reasonable rate. \cite{Arkles2013}
	}
    \label{fig:miscibility}
    \vspace{-5pt}
\end{figure}

The first class of material fabrication comes from a rapid and practical approach to manufacture large scale, homogeneous composites: emulsion templating.\cite{zhang2019}
In this case, one can mix liquid components together into an emulsion, pour the emulsion into a mould, and then polymerise (or otherwise solidify) one of the components to form the final product, which will typically consist of a SLC with generally spherical inclusions (e.g. \cite{Tutika2019liquid}, Figure \ref{fig:fabrication}a).
With the correct emulsion stabilisers, very high volume fractions of liquid can be achieved in the final SLC, even exceeding 74\% (random close packing for monodisperse spherical particles) for what are termed polymerised high-internal-phase emulsions or polyHIPE's.\cite{silverstein2014}
This type of material is particularly advantageous, as it can be produced at low temperatures, and easily formed -- for example, by the use of moulds or via 3D printing of the emulsion.\cite{neumann2020direct,Bartlett2017,lin2020attributes}

\subsubsection{Emulsion templating -- material constraints}

For these materials, a critical compatibility criterion is to choose material pairings that are not miscible, and do not react with each other, both in the emulsion state, and in the final state.
For liquids in hard, impermeable materials (e.g. in lubricant-impregnated materials), this is not normally a problem, as the liquids will not dissolve into a hard, inert phase.
However, liquid pairings, and soft elastomeric materials (primarily composed of lightly-crosslinked, polymeric liquids, potentially swollen with a liquid solvent) are more likely to be mutually soluble.
Thus for emulsion templating, finding immiscible materials pairings is particularly critical.

To get a sense for what sets miscibility, it is useful to briefly consider the thermodynamics of mixing.
This can be understood as a competition between entropy (which drives two components to mix with each other), and the enthalpy penalty that comes from mixing two dissimilar materials together.
Useful intuition for this comes from Flory-Huggins theory,\cite{flory1942thermodynamics} which describes the miscibility of two different chemical species.
This shows that miscibility is essentially controlled by two parameters: a mixing parameter, $\chi$, representing the relative strength of the enthalpic penalty to entropy, and the length of the polymer chains.
$\chi$ only depends on the nature of the physical interactions between the two materials, and the temperature, and is independent of the chain lengths of the two components.
Material combinations with large $\chi$ will be more immiscible.
Thus, the more dissimilar the materials (e.g. oil and water) and the lower the temperature, the less miscible the materials will typically be.
The length of the polymer chains is important as it is inversely proportional to the entropy of mixing.
Thus the longer the polymeric chains of the two components, the less miscible they will be.

Practically, this means that, in selecting materials, we can either choose liquids that are naturally immiscible ($\chi$ is large), such as a hydrocarbon and water, or we can choose liquids with a higher degree of polymerisation.
Fortunately, there is still a wide parameter space to work with.
Figure \ref{fig:miscibility} shows, broadly speaking, which of the six selected classes of materials are compatible with each other, in that they are not miscible, and do not degrade over time.
Note that in the figure, each of the material types can be thought of as a solid/liquid pair (e.g. water/hydrogel, silicone oil/silicone gel, ionic liquid/ionogel etc).
No lines between materials indicates that they have very low mutual solubility, and do not react with each other (e.g. silicones and LMs).
Dashed lines indicate that the materials are partially soluble, especially for low molecular weight liquids (e.g. fluorinated oils have a few percent solubility in silicones \cite{rosowski2020elastic}).
\edits{Lines} indicate that the materials are typically miscible (e.g. silicones and hydrocarbons).
Water and ionic liquids are a special case, as these can vary from immiscible to miscible, depending on the choice of ionic liquid.
Note that the figure is by no means exhaustive in terms of material combinations.
Liquid/solid pairs can even come from within the same class.
For example, two polymers dissolved in water will often phase separate, in a process termed liquid-liquid phase separation (e.g. \cite{shin2017}).
If one of these polymers forms a network, it will result in a phase-separated hydrogel/liquid composite.

\subsubsection{Emulsion templating -- fabrication}

Emulsion templating essentially consists of a mixing step to form droplets of one phase -- often by shearing -- followed by a solidification step (almost always polymerisation).
For this to be successful, the emulsion must be stable so that embedded droplets do not coalesce.
Typically, this requires the addition of sufficiently high concentrations of surfactant or small particles.
Both can produce composites with very high liquid volume fractions, and work by coating the interface between the two phases in the emulsion.
Interestingly, emulsions containing LMs often don't need a surfactant, likely due to a stabilizing, nanoscopic, metal-oxide layer that forms at the LM surface (e.g. \cite{Bartlett2016}).

At lower volume fractions, surfactant choice is not as crucial -- many surfactants will stabilize low concentration emulsions, and the only requirement is that the surfactant does not affect the polymerisation process (for example by partitioning out components of the polymerisation mix).
However, to achieve high volume fractions, surfactant choice is critical. This means choosing a surfactant that has two halves that each strongly absorb into the two phases of the emulsions (e.g. block copolymers of silicone and a hydrophilic polymer with the right geometry form excellent silicone/water emulsifiers).
Furthermore, the surfactant should follow Bancroft's rule, by being more soluble in the desired continuous phase of the emulsion (the final solid phase).
If it is not, the emulsion will invert, and the final product will be solid particles, stabilised in a continuous liquid phase.\cite{zhang2019}
Practically, the the suitability of common commercial surfactants and emulsifiers can often be determined from their reported HLB (hydrophilic/lipophilic balance), and further information from the manufacturer.

An effective alternative to using surfactants is the addition of nano- or micro-particles to form a Pickering emulsion.\cite{kataruka2019}
The particles adsorb strongly to liquid-liquid interfaces in the emulsion, where they then sterically prevent drop coalescence.
While surfactants often need to be used at relatively high loadings (often above 30\% of the prepolymer phase), Pickering emulsions can be stabilised with just a couple of percent of particles such as titania or silica nanoparticles, carbon nanotubes, or even clay.\cite{zhang2019}
The effectiveness of a given particle as a stabiliser chiefly depends on how it is wetted by the two liquid phases, with the best results when the particle are closer to neutrally wetting.
Particle shape,\cite{isa2014adsorption} stiffness,\cite{style2015adsorption} charge,\cite{zhang2019} and roughness \cite{zanini2017universal} can also all play an important role, both in how particles sit at interfaces, but also in how they adsorb to interfaces in the first place (e.g. \cite{albert2019}).

The polymerisation step can take one of multiple different routes.
For example, with silicone elastomers, the liquid prepolymer can be mixed with catalyst to trigger a slow polymerisation before the emulsion is made (e.g. \cite{style2015, somasundaran2006silicone, yilgor2014silicone, de2001silicone}).
For a chemically-crosslinked hydrogel (like polyacrylamide) that forms via free-radical polymerisation, polymerisation can be triggered by techniques such as directly adding a chemical initiator like ammonium persulphate (e.g. \cite{peng2019}), via UV illumination, or by raising the temperature in the presence of appropriate initiators.\cite{sheth2017uv, ma2017synthesis, gu2017mechanical} 
For a physically-crosslinked hydrogel, polymerisation can be controlled by factors like reducing the temperature (e.g. for agarose or gelatin) or addition of divalent ions (e.g. for alginate gels).\cite{zhou2018influence, cerciello2017synergistic}

\subsubsection{Emulsion templating -- controlling morphology}

The size, shape, and distribution of liquid inclusions that result from emulsion templating are controlled by the fabrication approach and ultimately determine many of the properties of the resulting SLCs (cf the Properties section later).
\edits{These properties can depend sensitively on these geometric parameters, especially as the inclusion size approaches the nano-scale, so they are important to control.
The geometry can be controlled over a broad size range}, as there are a range of different processes which can be used to form emulsions, broadly covered by shear mixing, sonication, acoustic waves, microfluidic/pipette approaches, and templating.
Furthermore, droplets can either be formed in situ in the elastomer matrix, or made in a solvent and then mixed into the uncrosslinked elastomer phase in a two-step approach.\cite{boley2015mechanically, Bartlett2016}

Droplets can be generated across a range of sizes.
Shear mixing generates emulsions with relatively larger droplets, with typical sizes in the 5-80 $\mu$m range.\cite{zhang2019,Tutika2018, Tutika2019liquid, Bartlett2016, tevis2014synthesis,kataruka2019}
Bath, or probe sonication results in droplets ranging between 10 nm and 5 $\mu$m.\cite{boley2015mechanically, hohman2011directing, lin2018sonication, farrell2018control, pan2019liquid, Tutika2019liquid}
Acoustic waves can create droplets either in the micro or nano range.\cite{tang2016chip, tang2019functional}
For each of these approaches, the droplet size distribution will be polydisperse. 
However, mono-disperse droplets can be generated drop-by-drop, typically using droplet dripping from openings of fixed geometry. \cite{tang2016liquid}
Normally, this is rather slow, but a number of microfluidic approaches have demonstrated how this can be done in a scaleable manner. \cite{amstad2016, dickey2008eutectic, tang2018microfluidic}

Liquid droplets typically form approximately spherical shapes due to surface tension in the absence of external forces, but there are a variety of techniques for forming non-spherical shapes.
In Ga-based LMs, a thin, but stiff oxide layer readily forms on the surface which can help to stabilize non-spherical shapes.
Thus, when ellipsoidal droplets are formed by flow focusing in a microfluidic channel filled with oxygen rich silicone oil, this oxide layer forms, and the droplets subsequently preserve their rice-grain-like shape when removed from the channel. \cite{hutter2012formation}
LM droplets formed in lower oxygen environments (e.g. water or deoxygenated silicone oil) tend to form microspheres. \cite{thelen2012study}
Rod-like solid particles have also been produced from Eutectic Gallium Indium (EGaIn) droplets via heating in an aqueous environment, which causes gallium to transform into crystalline gallium oxide monohydroxide, which takes a nano-rod morphology. \cite{lin2017shape}
Ellipsoidal inclusions can also be formed  by stretching the surrounding soft solid. \cite{Bartlett2017, Tutika2018, kazem2018}
This approach has recently been extended to form ellipsoidal LM inclusions in stress-free solids, using droplets in stretched and subsequently annealed thermoplastics. \cite{haque2020programmable}

In addition to size and shape, droplet distribution can also often be tuned.
Emulsification will typically give a uniform distribution of droplets throughout the sample, while concurrently setting droplet size.\cite{Tutika2019liquid} However, when the two phases in the composite have different densities, a settling step (potentially accelerated by centrifugation) can be used to create gradients in droplet density, yielding heterogeneous composite properties\cite{zhu2019anisotropic,neumann2020direct}
Settling can also be used to pack droplets together tightly, attaining volume fractions of filler that exceed the close-packing limits associated with solid composites (e.g. \cite{kataruka2019}).
Ultimately the length of the settling step depends on droplet size, viscosity, and density of liquid filler and matrix.\cite{wang2019highly}

\subsection{Phase separation \label{sec:phasesep}}
A second type of SLC are those that are formed by taking advantage of phase separation.
In this case, phase separation is triggered in a mixture of two components, for example by changing the temperature or changing a solvent, giving rise to droplet formation. 

A recent example of this made use of the temperature dependent solubility of fluorinated oil in silicone gels.\cite{style2018liquid} First, the gels are saturated with fluorinated oil at higher temperature (at a volume fraction of a few percent). 
When the gels are subsequently cooled, the solubility drops, and fluorinated oil phase separates, forming stable droplets embedded in the gel.
The size of the droplets are also fairly uniform and even tunable by the network cross-linking density, the cooling rate, and the composition of the solvent mixture. 

Similarly, any process that causes phase separation could be used to make composites.
For example, when water is added to concentrated anise oil/ethanol (ouzo) solutions, oil-rich droplets condense out as the more volatile ethanol evaporates,\cite{Tan8642} as shown in Figure \ref{fig:fabrication}b).
Similarly, when water is added to ouzo-soaked hydrogels, stable droplets condense inside the hydrogel.
When poly(vinyl alcohol) hydrogels are frozen, ice-filled pores are formed within the gel that remain as water pockets after the gel is thawed. \cite{mattiasson2009macroporous}
Other types of phase separation can be triggered by factors such as polymerisation of one component (recall miscibility typically drops significantly as materials polymerise, e.g. \cite{lee2010development}), or adding salt in many aqueous systems (e.g. \cite{thormann2012understanding}).

\subsubsection{Phase separation -- material constraints}

Due to the wide range of approaches to drive phase separation, it is not easy to give overarching rules for appropriate solid-liquid pairing.
However, the most important requirement, which differentiates it from emulsion templating, is that the components must be either partially or completely miscible before phase separation is triggered.
Partially-miscible solid/liquid pairings are a good place to start, as small changes (e.g. to temperature) will often significantly change solubility.
We also note that, when driving phase separation in solids, there is a significant barrier to nucleation and growth of droplets, which grows with increasing stiffness of the solid. \cite{rosowski2020elastic}
Thus softer solids can be easier to work with.

\subsection{Wetting and printing \label{sec:wetting}}

The final, general class of SLC is that where liquids are added to pre-formed solids.
This arises in many situations, including the printing of LM droplets onto soft surfaces to make soft electronic circuitry,\cite{boley2015mechanically} slippery lubricant-infused porous materials, \cite{lafuma2011,wong2011} and highly-stretchable, wicked membranes and fibres \cite{elettro2016,grandgeorge2018membrane} (see some examples in Figure \ref{fig:fabrication}c).

There are multiple different approaches to fabricating the solid, including soft lithography \cite{xia1998soft} and photolithography,\cite{schellenberger2015} spinning (to create fibres and membranes, e.g. \cite{huang2017preparation}), colloidal assembly, \cite{schaffner2015combining} subtractive manufacturing,\cite{ali2009fabrication} 3d printing, \cite{alison20193d} and adhesion of pre-made components.
These are too varied to cover in this review, so here we restrict ourselves to the discussion of some governing principles for ensuring correct wetting of solids, and of some recent liquid printing techniques.

\subsubsection{Wetting and printing
-- material constraints}

With these materials, it is typically important to control the wetting of the liquid on the pre-made solid.
Especially for the case of a lubricant-infused material, complete wetting is vital, as it allows the liquid to spontaneously wick into the porous solid, completely filling the pores, while minimising trapped air bubbles.
Furthermore, a wetting liquid will then stably adhere to the solid, preventing it from being removed by any external forces.

Fundamentally, wetting is determined by the surface tension of the liquid, $\gamma_{lv}$, and the surface energy of the solid/liquid and solid/vapour interfaces: respectively $\gamma_{sl}$ and $\gamma_{sv}$. \cite{RevModPhys.57.827}
When the spreading coefficient $S=\gamma_{sv}-\gamma_{lv}-\gamma_{sl}>0$, the liquid will totally wet the surface -- droplets will spread out into a film across the surface.
When $S<0$, droplets will not completely spread, and instead will have a finite contact angle, or partially wet the solid.
Thus, liquids with low $\gamma_{lv}$ will  typically spread better those with higher $\gamma_{lv}$.
Solid with higher $\gamma_{sv}$ will also promote wetting, as this increases $S$.
Furthermore, "like wets like": often when a solid and liquid are chemically similar, such as for a fluorinated oil and a fluorinated solid like teflon, $\gamma_{sv}\approx \gamma_{lv}$, and $\gamma_{sl}\approx 0$.
Then $S\approx 0$, and the liquid will wet the solid.

\begin{figure*}
 \centering
 \vspace{-5pt}
	\includegraphics[width=17cm]{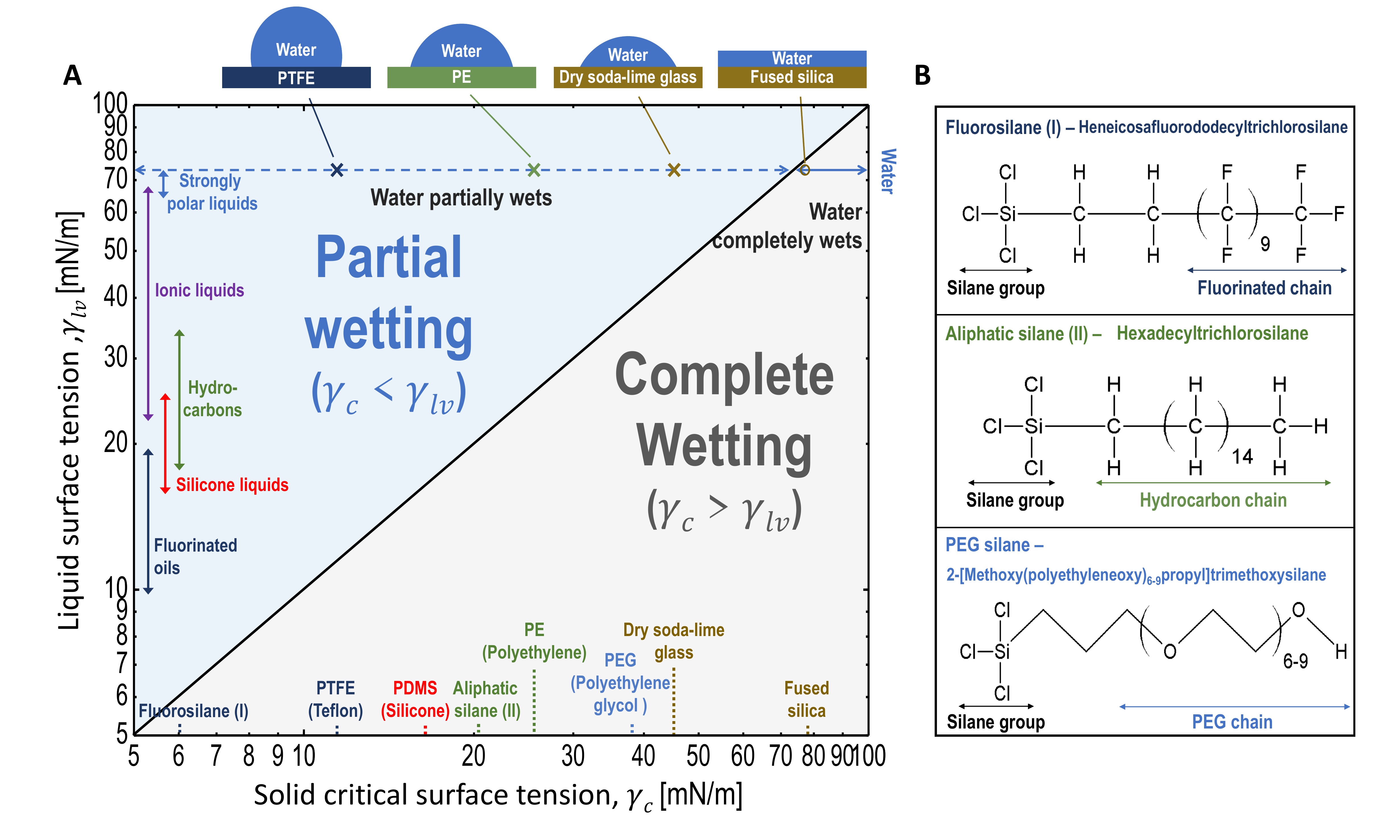}
    \vspace{-5pt}
	\caption{
	A) The different regimes of wetting behaviour for a range of liquids and solids. Complete wetting occurs when the critical surface tension of a substrate exceeds the surface tension of the wetting liquid. Values for various materials are taken from. \cite{Arkles2006,sedev2011surface,Arkles2013,SPEIGHT2017203}
	B) Silane coupling agents can coat a range of surfaces with polymer brushes of a variety of chemical forms, resulting in a modified $\gamma_c$. The table shows the three different silanes, two of which are referenced in A), which yield fluorinated, hydrocarbon, and hydrophilic brushes.
	}
	\label{fig:wetting}
    \vspace{-5pt}
\end{figure*}

Typically it is not necessary to calculate $S$.
Instead, it is observed that each surface is approximately characterised by a \emph{critical surface tension}, $\gamma_c$.
In general, liquids with $\gamma_{lv}<\gamma_c$ will completely wet the solid, while those with $\gamma_{lv}>\gamma_c$ will only partially wet it (Figure \ref{fig:wetting}).
The larger $\gamma_{lv}$ relative to $\gamma_c$, the higher the contact angle of the droplet.
Thus very few liquids will spread on teflon, with its low $\gamma_c=18.5$mN/m, while a range of liquids will spread on dry, soda-lime glass ($\gamma_c=47$mN/m).
Values of $\gamma_c$ have been tabulated for a wide variety of materials (e.g. \cite{arkles2014silane}), with a selection of common materials shown in the figure.

In the case that a desired liquid does not wet a certain solid, the value of $\gamma_c$ can also be  altered with a number of different surface treatments. \cite{otitoju2017superhydrophilic,li2016roles}
For example, silane coupling agents are commonly used to functionalise a range of surfaces, including glass, metal oxides, silicones and epoxy resin. \cite{plueddemann1991nature,zhou2012surface}
These molecules have one end that bonds to the solid surface, while the other end consists of a chain that is exposed on the functionalised surface, and thus determines its new wetting properties.
These chains can vary widely, with examples include hydrophilic, perfluorinated or hydrocarbon polymers (see Figure \ref{fig:wetting}b), to make the surface hydrophilic, fluorinated, or oleophilic respectively.
Each of these molecules can similarly characterised by a value of $\gamma_c$, as shown in Figure \ref{fig:wetting}a).
Other common surface modification techniques include corona treatment,\cite{sadeghi2013surface} chemical vapour deposition,\cite{jonsson1985chemical} physical vapour deposition,\cite{schneider2000recent} or grafting of polymer brushes.\cite{kostruba2013surface}
This means that a solid can be used that is convenient to fabricate complex structures, and then subsequently modified to have the correct wetting properties, allowing great control over the morphology of the final SLC.
\edits{Note, however, that these treatments are difficult to apply to the interior surfaces of porous solids.}

For the particular case of slippery, lubricant-impregnated surfaces (see the Properties section for more details), we note that it is useful to have a high degree of wetting specificity: the lubricant should wet the surface, but other liquids should not.
This stops the lubricant from being displaced from the solid when it is wet by another liquid.
In this case, a fluorinated surface, with $\gamma_c\sim 18$mN/m in combination with a fluorinated oil is a good combination.
Then, the low value of $\gamma_c$ means that most other liquids will not wet the substrate, while the fluorinated oil is not miscible with most other liquids.

\subsubsection{Wetting and printing -- fabrication}

One emerging application of SLCs is the fabrication of soft electronics via ink-jet printing of electronically-conducting liquids on soft substrates (sometimes followed by encapsulation).
Traditionally, the material has been metal/pigment particles in fluids but liquid fillers such as LM, and low melting point alloys are increasingly being utilized (e.g. Figure \ref{fig:fabrication}c).
For these, the typical process involves creating liquid-metal inks and print electronics by utilizing methods such as direct writing, stencil printing, and spray printing.
However, this is complicated for LMs, due to their high surface tension which makes the printing of small features below the capillary length challenging.
Instead, a promising technique for practical purposes involves the creation of micro- and nano- droplets in a carrier fluid and then dispensing them via established printing approaches.
Sonication and shear mixing have been reported as simple and rapid fabrication techniques to create tunable droplet sizes and concentrations.

After printing, the deposited liquid-metal droplets are typically insulating due to the oxide shell that forms on the surface.
Sintering through mechanical and thermal approaches breaks the oxide shell and merges the droplets to make them electrically conductive. Current strategies using this approach also involve injecting the solutions in microfluidic channels followed by solvent evaporation and mechanical sintering.\cite{lin2015handwritten,mohammed2017all}

\section{Properties of SLCs}

\begin{figure*}
 \centering
 \vspace{-5pt}
	\includegraphics[width=1\textwidth]{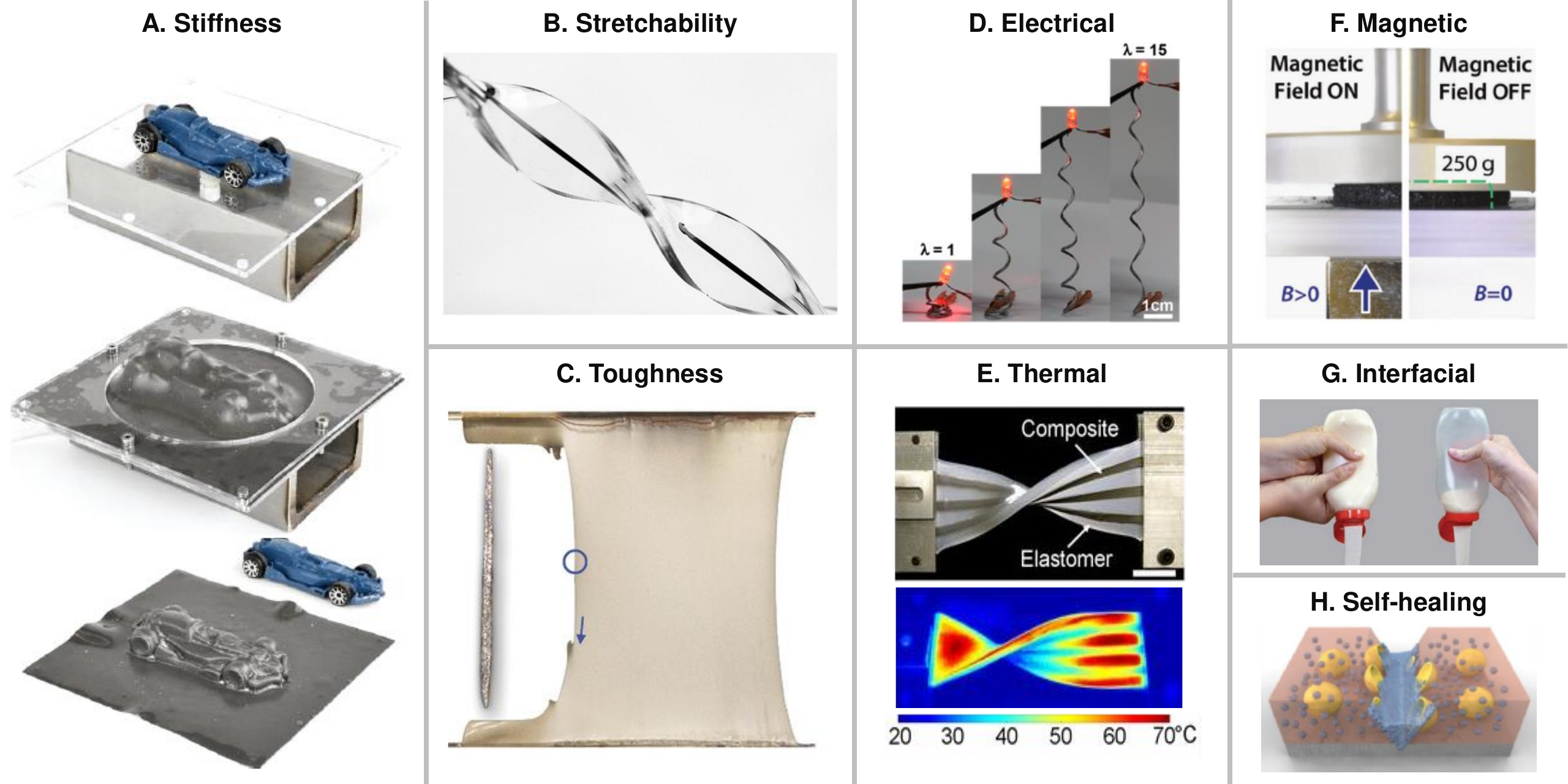}
    \vspace{-5pt}
	\caption{Enhanced properties through the addition of liquid inclusions. 
	A) A LM/silicone composite with temperature-dependant, tunable stiffness is used to take a mould of a 3D object by cooling the material after moulding. Reproduced with permission \cite{buckner2020shape} \copyright 2020, IEEE
	B) stretchable antennae \cite{Woo5088}. \copyright (2020) National Academy of Sciences
	C) liquid-metal/silicone composites show extremely toughness and crack blunting. Reproduced with permission\cite{kazem2018} \copyright2018, Wiley-VCH
	D) a conductor consisting of a liquid-metal/hydrogel composite\cite{park2019rewritable}. Reprinted with permission from \cite{park2019rewritable} \copyright (2019), American Chemical Society.
	E) enhancing thermal conductivity in soft materials with liquid-metal inclusions\cite{Bartlett2017}. \copyright (2017) National Academy of Sciences.
	F) the influence of the magnetic field on the stiffness of a polymer/magneto‐rheological fluid composite \cite{testa2019}. 
	Adapted with permission \copyright (2019), Wiley‐VCH.
	G) mayonnaise bottles with (left) no internal coating, (right) a lubricant-infused slippery surface coating. Image credit: LiquiGlide Inc. 
	H) schematic diagram of a typical self-healing process. Reproduced with permission \cite{cho2009self}. Adapted with permission \copyright (2009), Wiley‐VCH.}
    \label{fig:property}
    \vspace{-5pt}
\end{figure*}

\begin{figure*}
 \centering
 \vspace{-5pt}
	\includegraphics[width=\textwidth]{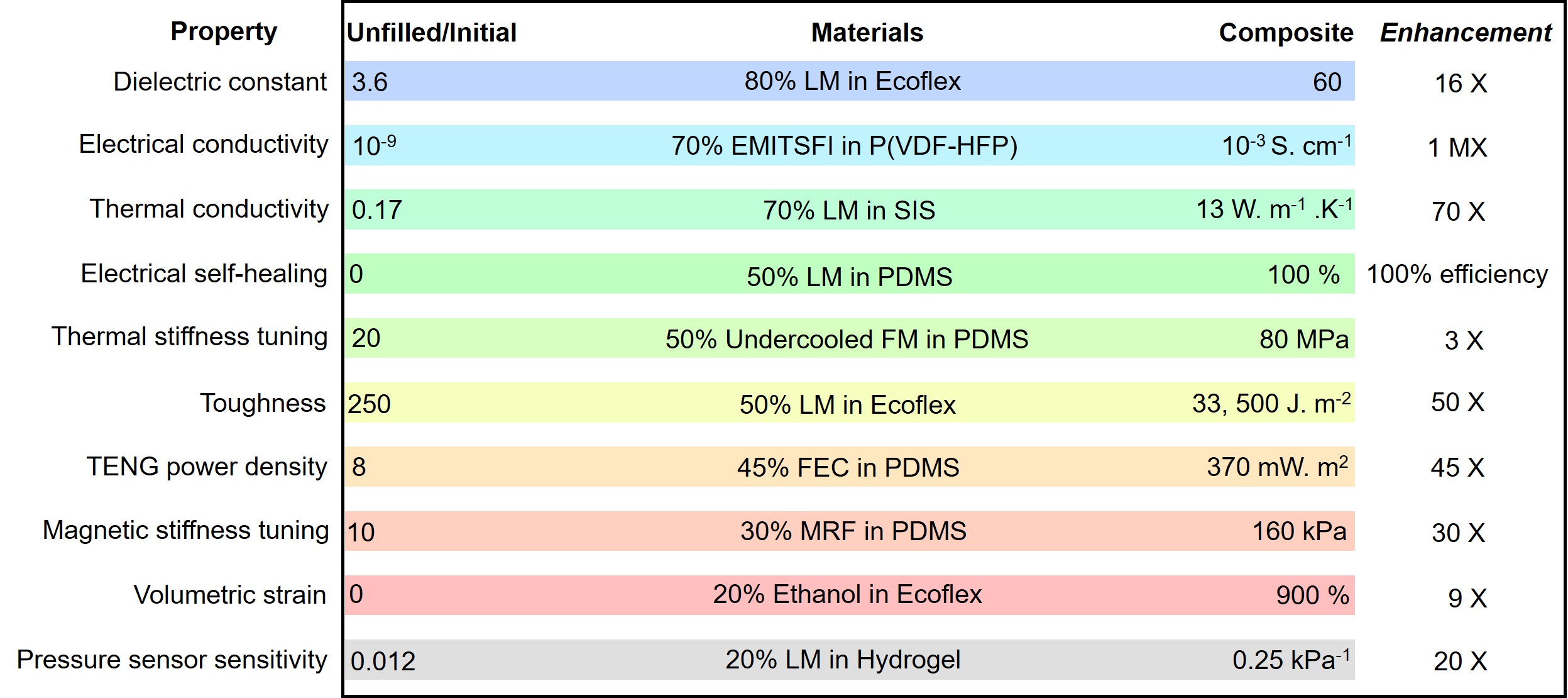}
    \vspace{-5pt}

\caption{Selected property enhancements using liquid fillers. \textbf{Dielectric constant} (LM droplets in ecoflex elastomer\cite{Tutika2019liquid}); \textbf{Electrical conductivity} (1-ethyl-3-methylimidazolium bis(trifluoromethylsulfonyl) imide
(EMITFSI) in poly(vinylidene
fluoride-co-hexafluoropropylene) P(VDF-HFP)\cite{cao2019self}); \textbf{Thermal conductivity} (programmed LM droplets in styrene-isoprene-styrene thermoplastic elastomer\cite{haque2020programmable}); \textbf{Electrical self-healing} (LM droplets in PDMS Sylgard 184\cite{Markvicka2018}); \textbf{Thermal stiffness tuning} (undercooled Field's metal droplets in PDMS Sylgard 184\cite{chang2018}); \textbf{Toughness} (LM droplets in ecoflex elastomer\cite{kazem2018}); \textbf{Tribolectric nanogenerator (TENG) power density} (fluoroethylene carbonate (FEC) in PDMS Sylgard 184\cite{jing2020liquid}); \textbf{Magnetic stiffness tuning} (magneto-rheological fluid (MRF) in PDMS from Gelest Inc.\cite{testa2019}); \textbf{Volumetric strain} (ethanol microdroplets in ecoflex elastomer\cite{miriyev2017soft}); \textbf{Pressure sensor sensitivity} (LM in polyacrylamide hydrogel\cite{peng2019})}
\label{fig:megaproperties}
\end{figure*}

SLCs can display a broad spectrum of different properties, depending on factors like their microstructure and composition, as illustrated in Figure \ref{fig:property}.
Their properties can show tremendous improvements over the properties of the pure matrix phase.
\edits{For example, a selection of such performance enhanced systems are presented in Figure \ref{fig:megaproperties} for properties ranging from electrical and thermal characteristics to mechanical properties and pressure sensor sensitivity. Although not exhaustive, the figure highlights the diverse properties that can be designed and provides a range of values as well as the scale of the potential enhancements for the respective properties.}
Here, we discuss some of the key properties that have been observed, including mechanical, electromagnetic, thermal and optical properties.
In some cases, depending on the composite microstructure, these can be predicted by rules of mixtures, and we outline these results, as these give insight into what range of properties can be achieved.

\subsection{Composite stiffness}\label{sec:compstiff}

\begin{figure}
 \centering
 \vspace{-5pt}
	\includegraphics[width=0.47\textwidth]{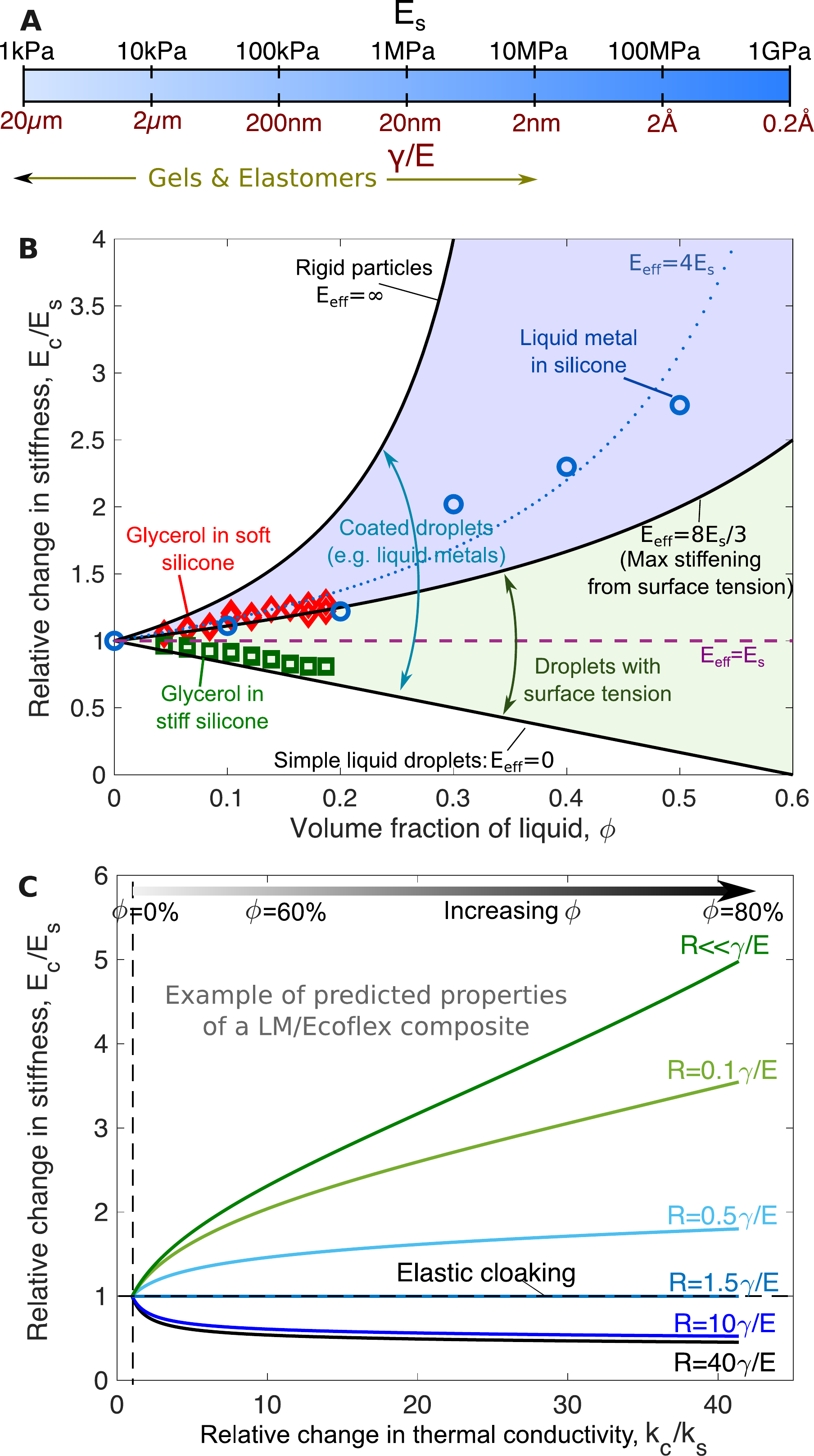}
    \vspace{-5pt}
	\caption{The effect of liquid inclusions on composite stiffness. A) Surface tension will stiffen a composite when liquid inclusions are smaller than the elastocapillary length. Here, we take a typical value of $\gamma=0.02$N/m to calculate this for a range of stiffnesses.
	B) Composite stiffness, $E_c$ as a function of the volume fraction of liquid inclusions. Red and green data points show relative change in $E_c$ as glycerol droplets are added to soft and stiff silicones respectively. \cite{style2015} Blue circles show the result of adding E-Gain to a silicone. \cite{Bartlett2016} Coloured areas show the range of accessible stiffnesses, following Eshelby theory. \cite{style2015} Droplets behave as elastic inclusions with stiffness $E_{\mathrm{eff}}$, that ranges from 0 (when there are no surface effects), to completely rigid (for LM droplets coated in a stiff oxide layer).
	C) The interplay between changes in mechanical and functional properties, illustrated via the change in normalised modulus as a function of normalised thermal conductivity as the volume fraction of LM filler is increased for different droplet sizes, $R$ relative to the elastocapillary length, $\gamma/E$.}
    \label{fig:stiffness}
    \vspace{-5pt}
\end{figure}

A good example of how liquids can change composite properties, and sometimes in surprising ways, is in their effect on a composite's stiffness.
In general, and especially for macroscopic liquid inclusions in solids, the addition of liquids will soften a composite, as elastic material is replaced by droplets with no resistance to shear.
However, when there are strong interfacial effects between the solid and liquid -- most commonly for microscopic inclusions -- then the liquid can actually stiffen the solid.
For example, when glycerol droplets are embedded in a soft silicone, the glycerol/silicone surface tension, $\gamma$ acts to keep the droplets spherical. \cite{style2015}
This effectively makes them more rigid, and thus their presence actually stiffens the composite (red diamonds in Figure \ref{fig:stiffness}).
When LM droplets are embedded in silicone, the Young's modulus of the resulting composite, $E_c$ can also increase with volume fraction, $\phi$ to actually be several times stiffer than the pure silicone, $E_s$ \cite{Bartlett2016} (blue circles in Figure \ref{fig:stiffness}).
In this case, this is due to the presence of a nanoscopic, stiff, oxide layer that forms at the LM/solid interface.
Again this opposes droplet deformation, effectively rigidifying the droplet, and thus stiffening the composite.

There are two key lengthscales that control the effect of liquid inclusions upon stiffness.
The first, the \emph{elastocapillary length}, $L_e=\gamma/E_s$ dictates when surface tension plays a significant role. \cite{style2017}
Only when inclusions are small relative to this, can surface tension play an important role often stiffening the composite.
The second pertains to LM droplets with thin, stiff interfaces.
In this case, behaviour is dictated by the ratio of inclusion size to the length $L_i=E_i h/E_s$, where $h$ is the thickness of the interfacial layer, and $E_i$ is its Young's modulus (or potentially its yield stress for weak layers).
When inclusions are smaller than $L_i$, then the interfacial layer is important, and the droplets' presence will stiffen the composite.

For the case of droplets embedded in solids, the effect of these two interfacial phenomena can be demonstrated using dilute Eshelby theory \cite{eshelby1957,palierne1990} (among other, more detailed models \cite{hill1965,mori1973,mancarella2016}).
This gives that, for a composite containing solid, spherical inclusions with Young's modulus $E_{\mathrm{eff}}$, the composite modulus is:
\begin{eqnarray}
    E_c=E_s\frac{1+\frac{2}{3}\frac{E_{\mathrm{eff}}}{E_s}}{\left(\frac{2}{3}-\frac{5\phi}{3}\right)\frac{E_{\mathrm{eff}}}{E_s}+\left(1+\frac{5}{3}\phi \right)}.
    \label{eqn:Ec}
\end{eqnarray}
Thus, we see that rigid inclusions have $E_c/E_s=(1-5\phi/2)^{-1}$, while liquid inclusions with no surface tension, and $E_{\mathrm{eff}}=0$ have $E_c/E_s=(1+5\phi/3)^{-1}$.
These are illustrated in Figure \ref{fig:stiffness}.

For droplets with a surface tension,  $E_{\mathrm{eff}}/E_s=24 L_e/(10 R + 9L_e)$, where $R$ is the droplet radius \cite{ducloue2014,style2015}, and both droplet and solid are assumed to be incompressible.
Thus $E_{\mathrm{eff}}$ can range from 0 up to $8E_s/3$ as $R$ decreases relative to $L_e$.
In particular, when $R\leq 1.5\gamma/E$, droplets effectively becomes stiffer than the surrounding solid, and thus increase $E_c$.
Inserting this into equation (\ref{eqn:Ec}) gives that capillary effects can cause stiffness changes with increasing volume fraction anywhere in the green area.

Equivalently for a stiff interfacial layer, we find that $E_{\mathrm{eff}}\sim E_i h/R$ (see Appendix).
In this case when $R \lesssim L_i$, we expect to see stiffening.
However, here there is no upper limit to the effective inclusion stiffness, so such droplets can result in composites with stiffnesses anywhere between the liquid, and rigid limits (i.e. the red and blue areas in Figure \ref{fig:stiffness}).

Note that this discussion assumes that composites consist of two entirely uniform phases.
However, this assumption should be used with caution.
For example, if a SLC is made by first preparing an emulsion, and then curing one of the components, often a stabilizer such as a surfactant or nanoparticles are required (e.g. \cite{chen2012,kataruka2019}).
In the case of nanoparticles, these will segregate to the liquid/solid interface, making it locally stiffer - in a similar manner to the oxide film.
Surfactants, on the other hand, can alter the curing process of the solid (e.g. \cite{style2015}), and this could change its resulting properties from those that are measured in bulk, especially in the neighbourhood of interfaces where surfactant concentrations are high.
It is also well established that many soft solids have properties that vary near interfaces -- especially hydrogels near hydrophobic surfaces \cite{gong2001} -- and this could result in substantially heterogeneous materials.

The stiffness of SLCs can also be actively tuned through phase transitions and applied magnetic or electric fields. These aspects are further discussed in \ref{sec:ETM_prop} and \ref{sec:App_SoRo} .

\subsection{Stretchability of composites}

One of the main benefits of working with SLCs in soft materials, is that the composite generally inherits the stretchability of the soft material.
Thus, working with solids like silicone gels and polyurethanes allows the fabrication of functional materials that can stretch by several times their original lengths (e.g. \cite{kazem2017soft,Markvicka2018}).
However, Nature has an elegant trick to achieve even greater extensibility with fibres and sheets (see Figure \ref{fig:types}).\cite{elettro2016}.
Certain fibres in spider webs have droplets of water on them that pull on the fibres with their surface tension.
If the fibre is loose, it will coil up inside the droplet until it reaches a critical tension - keeping a spider web naturally taut.
Because huge amounts of fibre can be spooled inside droplets, the fibre/droplet system can easily shrink and stretch drastically, while remaining tense.
This has been adapted to artificial materials with various polymeric fibre/liquid combinations to make highly stretching fibres and membranes, \cite{elettro2016,grandgeorge2018membrane} including conducting fibres that could stretch up to 20 times their original length. \cite{grandgeorge2018electricalfibre}

\subsection{Toughness \label{sec:tough}}

Unexpectedly, adding certain liquid inclusions has been shown to toughen soft materials \edits{ in response to static loadings}.
There are a variety of mechanisms to increase toughness in soft materials which operate primarily through  the deflection, and blunting of crack tips and increased dissipation near the crack tip. \cite{creton2016fracture, zhao2014multi,gong2003double, lake1967strength} Liquid inclusions can contribute to both of them. 
For example, micron-scale LM inclusions dispersed in soft silicone elastomers elongate in front of a crack tip in the material.\cite{kazem2018}
The crack tip appears to avoid the droplets, so the elongated droplets cause the crack to deflect perpendicular to its original path (Figure \ref{fig:property}C).
Thus, the crack ends up travelling parallel to the applied load.
This `knotty tearing’ is similar to behaviour seen with some rigid filler/elastomer combinations like carbon black and natural rubber. \cite{hamed2003effect, de1996tear}
In the LM case, it leads to a 50x increase in fracture toughness while maintaining a soft, extensible material.
Dissipation can be increased when adding either magnetorheological fluids, \cite{testa2019} or LMs to elastomers — both for microscopic, \cite{kazem2018} and macroscopic droplets. \cite{owuor2017nature}
This shows further promise for toughening and control of kinetic energy in soft materials, \edits{especially in response to dynamic loadings.
For example, yield-stress inclusions have been used to increase the penetration resistance of materials to create `liquid armor'. \cite{lee2003ballistic}
Along similar lines, MRF inclusions have been used to make impact-absorbing materials with rapidly-switchable levels of energy absorption. \cite{deshmukh2006adaptive}}

\subsection{Electrical, thermal, and magnetic properties}\label{sec:ETM_prop}

Soft materials typically have poor electrical and thermal conductivities, and little magnetic responsiveness.
Liquid fillers can improve these, while maintaining the desirable soft, extensible properties of the material \cite{dickey2017stretchable} (e.g. Figure \ref{fig:property}B)-- unlike the case for traditional rigid inclusions. \cite{Mamunya2002a, Yang2011, Kim2006,Hong2019,Ghose2006HPP, Yang2011}

For improving thermal properties, LMs are desirable due to their high thermal conductivities relative to other liquids (see section 3.1)  \cite{Bartlett2017, Jeong2015, Ralphs2018, Tutika2018, Zhao2018a, Fan2018,haque2020programmable} (c.f. Figure \ref{fig:property}D).
In particular, EGaIn is is often used, as it is liquid at room temperature, \cite{kazem2017soft} and has low viscosity \cite{Khondoker2016} and low toxicity. \cite{Chiechi2008ACIE}
For example, the thermal conductivity of EGaIn droplet/silicone composites has been shown to increase thermal conductivity from 0.2 W/mK (for the base elastomer) by 5 - 70x. \cite{Bartlett2017,Jeong2015, Tutika2018, haque2020programmable}
This enhancement of thermal conductivity is often a result of the unique thermal-mechanical coupling of SLCs. The elastic nature of the matrix accommodates high strains which can shape the LM droplets into aligned ellipsoids with a high aspect ratio in the stretching direction. This leads to thermally conductive pathways which can display metal-like thermal conductivity while still remaining flexible and conformal.\cite{haque2020programmable} 
There are a variety of effective medium theory (EMT) approaches  to theoretically predict the thermal properties of SLCs.
As an example, recent work has shown that, for ellipsoidal droplets, Bruggeman theory matches well with experimental results\cite{bruggeman1935berechnung}:
\begin{eqnarray}
\label{eqn:Bruggmann_Eqn}
    \bigg(\frac{k_p - k_c}{k_p - k_s}\bigg) \bigg(\frac{k_s}{k_c}\bigg)^L = 1-\phi
\end{eqnarray}
where $k_p$, $k_s$, and $k_c$ represent the thermal conductivity of the embedded droplets, solid matrix, and composite, respectively.  The depolarization factor $L$ represents the droplet shape, and reduces to 1/3 when they are spherical.

LMs are also utilized for electrical applications, due to their huge (relative to other options) electrical conductivity and dielectric properties \edits{(e.g. \cite{Bartlett2016,Koh2018a, xin2019ultrauniform, ford2019multifunctional,bartlett2019self,park2019rewritable}, Figure \ref{fig:property}D)}.
In general, most as-fabricated, discrete liquid in continuous solid liquid-metal composites are insulating, both due to the insulating elastomer phase, and due to insulating oxide layers around the LM droplets. However, connected LM pathways can be introduced into soft solids, to form highly stretchable conductors.
For example, this has been done by encapsulating films of LM in between two elastomer layers, \cite{boley2015mechanically,mohammed2017all,lin2015handwritten,Thrasher2019} by filling porous foams with LMs, \cite{liang2017}, \edits{filling frozen LM foams with curing elastomer, \cite{yu2020super,yao2020highly}} and by injecting LMs into microfluidic channels. \cite{kubo2010stretchable,chossat2013soft,frutiger2015capacitive,dickey2017stretchable,palleau2013self,li2016galinstan}
Electrically connected pathways in composites containing dense packings of droplets can also be created by connecting the droplets together.
This is achieved through localized pressure (mechanical sintering), light (laser sintering), chemicals (chemical sintering), and temperature can rupture the droplets resulting in pathway formation. \cite{lin2015handwritten, boley2015mechanically,martin2019heat, wang2019highly}
Soft elastomer/LM composites have even been shown to have self-healing abilities, as the LM droplets maintain connectivity in the material, even after electrical pathways are cut. \cite{Markvicka2018}

Beyond stretchable conductors, LMs can also be used to create insulating materials with high dielectric constant (equivalently high permittivity) with a range of uses including soft actuators, generators, sensors and loudspeakers.\cite{wang2020soft}
This takes advantage of the naturally insulating properties of un-connected LM droplets in soft materials.
For example, recent work has shown that the dielectric constant of a filled rubber can be increased up to $\sim$ 40x that of its unfilled counterpart, taking values of 70-80 with droplet loadings of up to 80\%\cite{Tutika2019liquid, Koh2018a}
It is worth noting that other liquids can also be utilised — for example fluorinated oil droplets have also been dispersed in silicone elastomers to increase dielectric constant and enhance the performance of triboelectric nanogenerators (TENG) (with a 45X improvement in power density). \cite{jing2020liquid}
The change in permittivity can also be described by EMT theories, such as analytical theories by Nan et al.\cite{nan1997effective,Tutika2019liquid, Bartlett2016}

In terms of magnetic responsiveness, MRFs dispersed in soft elastomers have been shown to give rise to a range of magnetically-tunetable properties.\cite{York2007,Bastola2018} For example, soft silicone gels containing dispersions of MRF droplets had Young's moduli that could be increased up to the 30x by the application of a magnetic field, while also displaying shape-memory behaviour with shape transition times of less than a second. \cite{testa2019}
Additionally MRF/elastomer composites have been shown to yield magnetically controllable dissipative properties, due to the strongly field-dependent loss properties of the MRF.
Thus both loss tangent \cite{testa2019} and loss modulus \cite{bastola2017novel} have been shown to be controllable via magnetic fields — each changing by a factor of 2-3x in the presence of a field.
MRF has been selectively deposited into the the elastomer by 3D printing, to create well-defined structures of MRF, rather than dispersed droplets. \cite{bastola2017novel}
We note that liquid-metal based MRFs do exist, \cite{hu2019,elbourne2020antibacterial,merhebi2020magnetic} so although we have not seen their use in SLCs, it should also be possible to create materials that have simultaneously improved electric, thermal and magnetic properties.
Such liquid-metal/iron particle mixtures have even shown novel properties such as magnetically-switchable antibacterial behaviour, so pose interesting possibilities for future material combinations.

Further hybrid or multiphase particle droplet systems have been demonstrated which can increase performance and enable multifunctionality.
For example, these hybrid systems can be utilized to enhance the thermal and electrical properties of LM systems through the addition of metal films or particles such as copper, silver, nickel, and iron.\cite{hirsch2016intrinsically,tang2017gallium,guo2018ni}
For example, tungsten can increase the thermal conductivity of LM by a factor of 3x.\cite{, kong2019oxide}
Silver particles can also be added into eGaIn to increase electrical conductivity for printed electronics.\cite{tavakoli2018egain} 
Solid particles can also modify the viscosity of LM, \cite{nesaei2019rheology} and depending on the combination of solid and liquid particle materials chosen, the materials can alloy together forming intermetallic compounds, which can change the liquid nature of the filler into a solid or semi-solid mixture.\cite{Ralphs2018, Tutika2018, tang2017gallium, tavakoli2018egain, yun2019liquid}

\subsection{Optical properties}

\begin{figure*}
 \centering
 \vspace{-5pt}
	\includegraphics[width=17cm]{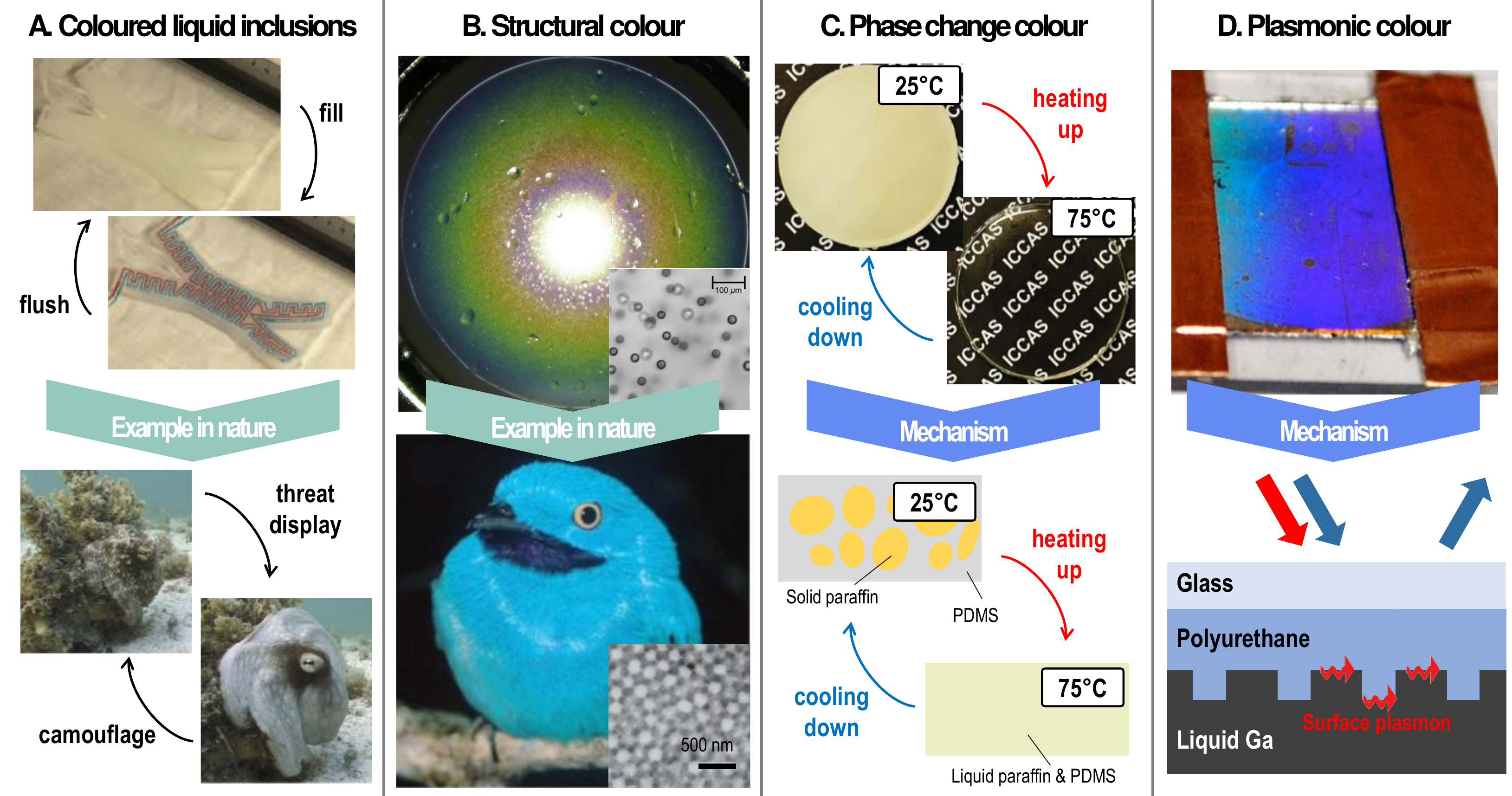}
    \vspace{-5pt}
	\caption{Examples of SLCs with changed optical properties.
	A) Coloured liquid inclusions in artificial. From \cite{morin2012}, reprinted with permission from AAAS. Reprinted from  \cite{hanlon2007cephalopod}, \copyright (2007), with permission from Wiley-VCH. 
	B) Structural colour from a fluorinated oil/silicone composite, \cite{style2018liquid} and in nature \cite{dufresne2009self}. Reproduced from Ref. \cite{dufresne2009self} with permission from The Royal Society of Chemistry.
	C) Switchable colour driven by phase change. Reproduced with permission \cite{yao2014} \copyright2014, Wiley-VCH.
	D) Switchable plasmonic colour that changes upon melting/solidificaiton of the LM layer. Reprinted with permission from \cite{vivekchand2012} \copyright (2012) American Chemical Society.}
    \label{fig:optical}
    \vspace{-5pt}
\end{figure*}

Adding liquids to clear solids offers multiple routes to change their optical properties, as illustrated in Figure \ref{fig:optical}.
Most simply, a dye can be added to the liquid to colour the composite (Figure \ref{fig:optical}A).
If the liquid can be exchanged, this allows an easy approach to dynamic colouring of the material. \cite{morin2012}
Alternatively, cephalopods stretch out small pockets of dye to rapidly change their surface colour. \cite{hanlon2007cephalopod}
Without dye, composites containing small, polydisperse liquid inclusions will typically appear white (like milk), due to strong scattering of light off all the surfaces in the composite.

However, dye is not essential for colour production.
For example many animals, including various butterflies and birds, use \emph{structural colour} to create vibrant hues -- typically blues and greens. \cite{dufresne2009self}
The key principle here is that colours start to emerge when composite inclusions are monodisperse, with a size comparable to the wavelength of light.
In this case, certain wavelengths will be scattered strongly by the inclusions and others will travel through the composite relatively unaffected.
Then the strongly reflected colour is seen as colour by an observer.
The effect can be reinforced by having the inclusions in certain structural arrangements, like in a crystalline array, or randomly arranged with a well-defined average spacing.
Recent work has shown that such colour can be produced in a liquid composite via phase-separation-driven formation of droplets in soft solids \cite{style2018liquid} (Figure \ref{fig:optical}B).

There are also a number of approaches to creating tunable colour with SLCs.
As one example, paraffin-soaked gels are transparent when the paraffin is liquid, but go opaque at its freezing point \cite{yao2014} (Figure \ref{fig:optical}C).
A completely different idea uses LM surface-plasmon resonances (Figure \ref{fig:optical}D).
In this case, surface plasmon polaritons (surface electron resonances) at the interface of liquid-metal inclusions in a structured solid give rise to vivid colour generation.
When the LM freezes, its plasmonic surface properties change significantly, and the colour is lost. \cite{vivekchand2012}

\subsection{Interfacial properties}

Not only can adding liquid components to solids change their bulk properties, they can also significantly alter surface interactions like wetting and adhesion.
One example of controlling interfacial interactions is a lubricant-impregnated surface, inspired by the strategy carnivorous pitcher plants use to trap prey. \cite{bohn2004}
This consists of a microporous, or textured solid, and a liquid lubricant that wets it.
The liquid then stably coats the surface, and acts as a lubricating layer, so that objects on the surface slide off easily. \cite{lafuma2011,wong2011,schellenberger2015}
In the case of the pitcher plant, ants cannot grip their slippery rim, and fall into the pitcher, where they are digested.
Demonstrated applications include anti-icing, \cite{kim2012} anti-fouling, \cite{tesler2015} non-stick coatings, \cite{lafuma2011} and enhancing condensation rates. \cite{anand2012}
For these slippery surfaces, the adhesive properties depend heavily on the solid/liquid combination chosen, but is generally most effective with liquids that have low surface tensions such as silicone oils, fluorcarbons (e.g. Krytox\texttrademark), or hydrocarbons. \cite{solomon2016}
As such a broad range of different materials can be used, they can be easily tailored to different applications.
However, it also gives rise to the main problem with such surfaces: these liquids are often volatile, which can leads to depletion of the lubricant, and thus degrade their properties over time (though there are now a number of strategies available to mitigate this effect. \cite{solomon2016})

As the liquid properties typically control adhesion, it makes sense that the use of tunable liquids can result in tunable adhesion.
For example, using a liquid that solidifies for a lubricant-impregnanted surface gives rise to drastically changing wetting and adhesion at the liquid's melting point. \cite{yao2014}
More fine control can be achieved with ferrofluids and MRF's, which can be displaced, and change their rheology upon magnetisation.
For example, Wang et al. created a ferrofluid-infused structured surface, where the ferrofluid was able to drain down through the texture on application of a magnetic field. \cite{wang2018}
This caused the surface to change from smooth and wet, to rough and dry upon magnetisation, and hence from slippery to sticky. Testa et al. \cite{testa2020switchable} have dispersed MRF droplets in a soft, silicone gel.
The composite increases adhesion (i.e. pull-off force) when it is magnetised, as the MRF becomes a yield-stress solid, and dissipates a significant amount of energy as an indenter is pulled off the surface.

\subsection{Self-Healing}
SLCs have been used to create self healing systems.
Self-healing can be designed both for repairing physical damage, and for repairing electrical connections.\cite{bartlett2019self}
For the case of SLCs, most approaches have focused on either droplets of healing agent or structured, vascular networks.\cite{patrick2016polymers} 
Molecular mechanisms such as dynamic bonds have also been used, but these approaches are out of scope for our discussion.\cite{burnworth2011optically,yang2013self}

For capsules or structured networks, damage dispenses the healing agent into the damage zones. For the case of material self-healing, the liquid phase is typically a monomer with a catalyst that will react with the polymer matrix,\cite{white2001autonomic,toohey2007self} or multiple capsule systems can also be used which contain different chemical functionalities.\cite{keller2007self,cho2009self}   
For electrical self-healing systems, the liquid phase is typically a LM or a fluid phase filled with conductive particles.\cite{palleau2013self,Markvicka2018, blaiszik2012autonomic,odom2012autonomic,chu2018smart,li2016galinstan,xin2019ultrauniform,guo2019magnetic}
The time scale and efficiency of the healing are key metrics for the performance of self-healing systems. 
For material self-healing the time scale is often dictated by the chemical reaction to create new bonds, and can range from 100s of seconds to days. For electrical self-healing, LM networks have been shown to self-heal instantaneously as LM droplets reconfigure during the damage event.\cite{Markvicka2018}
For efficiency, material healing has been shown to be commonly greater than 60\% with examples of up to 100\%.
For electronics, efficiency is commonly greater than 80\%, while efficiencies can effectively go above 100\% if the circuit physically reconfigures to resistors in parallel instead of a single conductive trace in series.\cite{park2019stretchable,Markvicka2018}    

\subsection{Changing multiple properties simultaneously}
One interesting opportunity with SLCs is the ability to independently control multiple properties (e.g. stiffness and thermal conductivity).
In the case of discrete liquid droplets in a continuous solid, the properties are typically dictated by three parameters: the choice of liquid filler (c.f. section \ref{sec:liquidchoice}), loading ($\phi$), and the size of the droplets ($R$).
If the desired properties depend differently on $R$ and $\phi$, they can be independently tuned.
To demonstrate this, in Figure \ref{fig:stiffness}C, we show a typical plot of how elastic modulus ($E_{c}/E_{s}$) and thermal conductivity ($k_{c}/k_{s}$) change in a LM-silicone elastomer system with $\phi$ and $R$.
These are calculated using Equations (\ref{eqn:Ec},\ref{eqn:Bruggmann_Eqn}), assuming that the LM droplets behave as simple liquids with a surface tension (i.e. no oxide layer).
Thermal conductivity only depends on loading, so can be selected simply by varying $\phi$.
The modulus depends on both $R$ and $\phi$, so it can be subsequently tuned by varying the droplet size relative to $\gamma/E$.
Specifically, for a given $k_{c}/k_{s}$ as $R$ becomes much less than $\gamma/E$, the normalized elastic modulus increases, while for $R$ greater than $\gamma/E$ the modulus decreases. 
A special ideal case at $R = 1.5\gamma/E$ is observed, where the droplet is "cloaked" and the liquid filler shows negligible stiffening or no change in the normalized elastic modulus while thermal conductivity is increased. 
This representative plot provides an example of the relevance of choosing materials and fabrication approaches (i.e. to set particle size) to achieve specific mechanical and functional properties/performance.

\section{Applications}
\begin{figure*}
 \centering
 \vspace{-5pt}
	\includegraphics[width=1\textwidth]{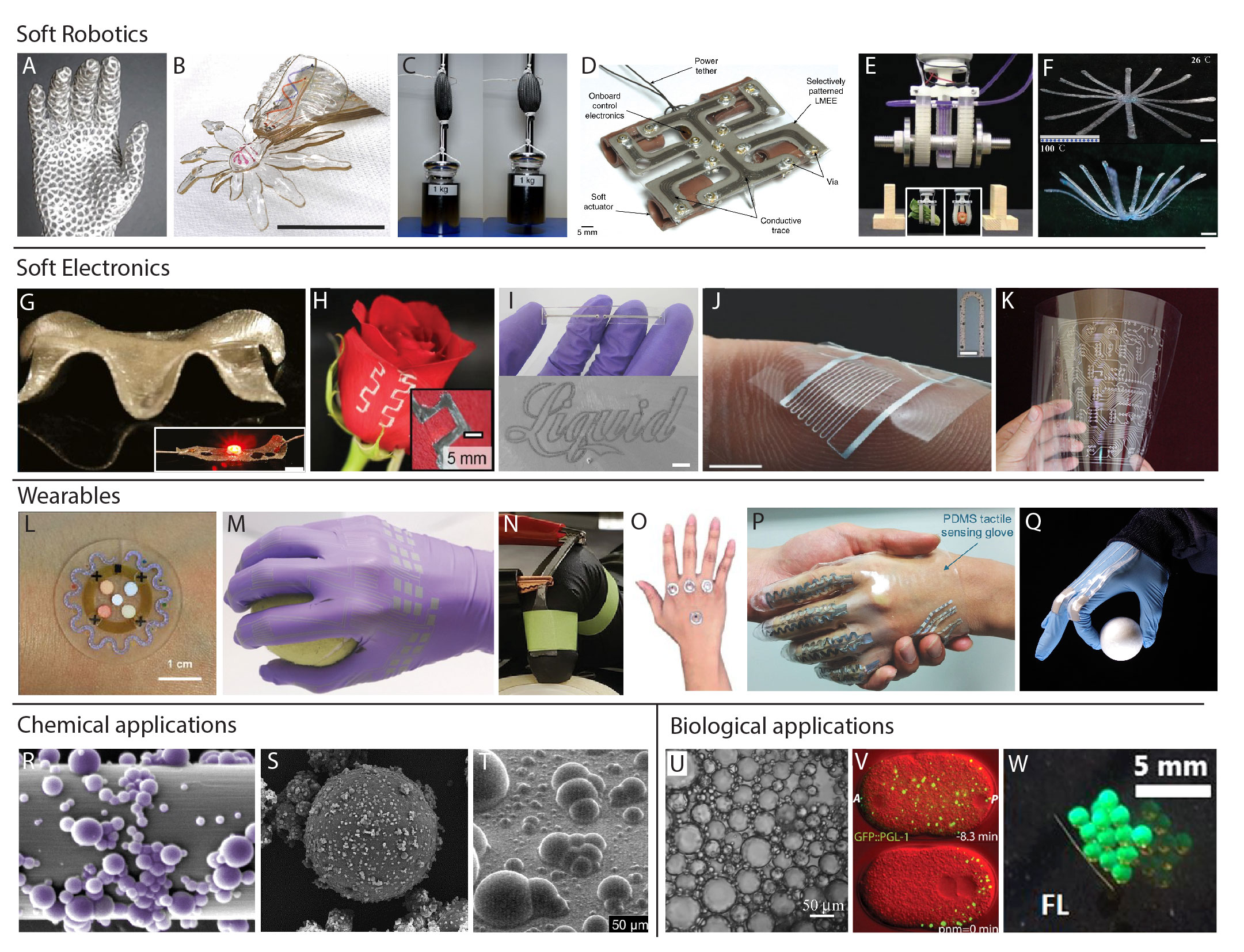}
    \vspace{-5pt}
	\caption{Application examples. 
\emph{Soft Robotics} 
	A) A multifunctional LM lattice with a shape memory effect and tunable shape and rigidity. Reproduced with permission\cite{deng2020multifunctional} \copyright2020, Wiley-VCH. 
	B) A soft robotic peacock spider with actuatable appendages. Reproduced with permission\cite{ranzani2018increasing} \copyright2018, Wiley-VCH. 
	C) A McKibben-type soft artificial muscle. Reproduced with permission.\cite{miriyev2017soft} \copyright 2017, Springer.
	D) A soft, quadruped robot, connected with autonomously self-healing conductive traces made from LM droplets. Reproduced with permission.\cite{Markvicka2018} \copyright 2018, Springer.
	E) A soft actuator made by multimaterial 3D printing. Reproduced wih permission\cite{zhang2019fast} \copyright2019, Wiley-VCH.
	F) A temperature-responsive, deforming bilayer material, that works based on the presence of embedded liquid droplets. Reproduced with permission\cite{li2019multimaterial} \copyright2019, Wiley-VCH.;
\emph{Soft Electronics} 
	G) Shape-morphing LM-LCE composites.\cite{ford2019multifunctional} \copyright (2019), National Academy of Sciences. 
	H) Electronic circuitry on a delicate, rose petal, created with heat-free metallic electrical interconnects. Reproduced with permission\cite{martin2019heat} \copyright2019, Wiley-VCH. 
	I) Soft circuit boards created with LM nanodroplets. Reproduced with permission\cite{lin2015handwritten} \copyright2015, Wiley-VCH. 
	J) Stretchable, biphasic thin metal films. Reproduced with permission\cite{hirsch2016intrinsically} \copyright2016, Wiley-VCH. 
	K) Personal electronics printed with a composite LM ink. Reproduced with permission.\cite{zheng2014personal} \copyright 2014, Springer.; 
\emph{Wearables} 
	L) Soft, wearable microfluidic device for sweat sensing. From\cite{koh2016soft}, reprinted with permission from AAAS.
	M) A functionalised glove inkjet-printed with EGaIn droplets. Reproduced with permission\cite{boley2015mechanically} \copyright2015, Wiley-VCH. 
	N) Stretchable resistive heaters through polymerised LM networks. Reproduced with permission\cite{Thrasher2019} \copyright2019, Wiley-VCH. 
	O) Liquid-metal antennae for wireless human motion monitoring. Reproduced with permission.\cite{jeong2017skin} \copyright 2017, Springer.
	P) A wearable microfluidic pressure sensor for tactile touch monitoring. Reproduced with permission\cite{gao2017wearable} \copyright2017, Wiley-VCH. 
	Q) All-soft-matter gesture monitoring glove.\cite{Tutika2019liquid} Reprinted with permission from \copyright (2019) American Chemical Society.; 
\emph{Chemical applications} 
	R) \edits{Autonomic healing at carbon fiber/epoxy interfaces functionalized with capsules containing reactive epoxy resin and ethyl phenyl acetate (EPA).\cite{jones2014autonomic}  Reprinted with permission from \copyright (2014) American Chemical Society.}
	S) Epoxy capsules with a urea-formaldehyde shell for self-healing composites. Reproduced with permission\cite{yin2007} \copyright2017, Wiley-VCH. 
	T) Frost formation on lubricant-impregnated surfaces.\cite{rykaczewski2013mechanism} Reprinted with permission from \copyright (2013) American Chemical Society.; 
	\emph{Biological applications} 
	U) \edits{Cinnamon oil-loaded composite emulsion hydrogels with excellent antibacterial properties. Reproduced with permission\cite{wang2018cinnamon} \copyright2018, Wiley-VCH.}
	V) P-granules: protein droplets (green) in a \emph{C. elegans} embryo. From\cite{brangwynne2009germline}, reprinted with permission from AAAS. 
	W) Droplets containing green fluorescent protein embedded in silicone. Reproduced with permission.\cite{faccio2019} \copyright 2019, Springer.}
    \label{fig:applications}
    \vspace{-5pt}
\end{figure*}

As we have seen, the combination of soft mechanical response with tunable functionality in SLCs produces a broad range of materials with useful properties. Due to their versatility, they find utility in a wide variety of applications ranging across the fields of soft matter and engineering, as demonstrated in Figure \ref{fig:applications}.  
Here, we discuss some of these applications in soft robotics, soft electronics, and chemical and biological systems.

\subsection{Soft robotics}\label{sec:App_SoRo}
Soft robotics, based on the use of compliant components, has arisen recently as an excellent way to make human compatible robotics, which can perform functions not currently accessible in rigid robotic systems.\cite{rus2015design,majidi2014soft}  
To this end, SLCs can provide functional components that can be tailored for actuation, sensing, rigidity tuning, and computing. \cite{wehner2016integrated}
Several examples are shown in Figure \ref{fig:applications}A-F:
Figure \ref{fig:applications}A shows a LM lattice material with a shape memory effect achieved by harnessing the solid-liquid phase transition of LMs to achieve tunable rigidity.
Figure \ref{fig:applications}B shows an actuatable, soft-robotic peacock spider with embedded microfluidic circuitry that is manufactured using multilayer soft lithography (here 12 layers), laser micromachining, and folding to create 3D soft microstructures and devices. 
Figure \ref{fig:applications}E shows a fast‐response, stiffness‐tunable (FRST) soft actuator, that is pneumatically driven, and reduces its stiffness with temperature, so that it can be kept in a stiff, glassy state when not being actuated.
For this purpose, the FRST contains printed joule-heating circuit and microfluidic cooling channels that allow it to complete a softening–stiffening cycle within 32 s, all fabricated via a hybrid multimedia 3D printing approach.
Finally, Figure \ref{fig:applications}F shows a 3D printed material consisting of patterned water/glycerol droplets in a resin matrix, fabricated by combining direct ink writing and microfluidics.
When heated up, the object undergoes a differential thermal expansion,  controlled by the droplet microstructure, which allows it to change shape.

\subsubsection{Actuation}
Actuation in  liquid filler systems can be achieved with a variety of methods.
For example, large strains can be achieved with a SLC consisting of ethanol microdroplets dispersed in an elastic silicone rubber.
When a current is supplied to a thin resistive wire, the temperature of the ethanol reaches its boiling point of 78 $^\circ$C, it boils and the vapors result in a significant volumetric expansion of 900\%. \cite{miriyev2017soft,li2020soft}
When implemented as a bicep, the actuator mimics the radial expansion and longitudinal contraction of a natural muscle (Figure \ref{fig:applications}C).
Volatile liquids like this may slowly leech out of elastomeric composites, often limiting lifetime to such materials.
More durable, but less extreme results can be achieved with inclusions of a less volatile material like paraffin, which changes its volume by around 10-15\% upon melting. \cite{sharma2011viton,ogden2014review}
A second, recently-demonstrated technique involves hydraulically amplified self-healing electrostatic (HASEL) actuators, consisting of a liquid-filled polymeric pouch which shows rapid and large deformation upon the application of an electric field (e.g. Figure \ref{fig:types}D). \cite{acome2018hydraulically,kellaris2018peano}
Here, stretchable electrodes are placed on either side of a dielectric membrane. Upon the application of large voltages, the Maxwell stress pulls the outer surfaces together which also produces deformations outside of the active, electrode area which can be leveraged to do rapid work.
Further approaches include actuation via electrohydrodynamic pumping of liquid within elastomeric channels, \cite{cacucciolo2019stretchable} embedding of LM droplets into liquid crystal elastomers (LCEs) to enable electrically activated shape-shifting structures (Figure \ref{fig:applications}G), \cite{ford2019multifunctional} and encapsulating electrically-controlled, shape-memory-alloy (SMA) wires in LM/elastomer composites.
In this last case, the composite material serves to give low mechanical resistance to actuation which enhances the actuation frequency and deflection, while dissipating heat generated by the SMA wires. \cite{Bartlett2017}
LM droplets can also be embedded into liquid crystal elastomers (LCEs) to enable shape shifting structures which can be activated electrically (Figure \ref{fig:applications}G).\cite{ford2019multifunctional}

\subsubsection{Sensing}
Sensing in SLCs can detect deformation, state, or damage.
Commonly, these sensors are composted of microfluidic channels of electrically conductive liquid such as LMs,\cite{park2010hyperelastic,kramer2011wearable} however other fluids such as ionic liquids and conductive grease can also be used.\cite{muth2014embedded,choi2017highly,chossat2013soft}
For detecting deformation, resistive and capacitive sensing are often used.\edits{\cite{polygerinos2017soft,chen2018transparent,yang2020ultrasoft}}   In these scenarios the deformation changes the geometry of the sensing material which gives rise to a measurable change in signal with a behaviour that can be straightforwardly modelled.
For example, for a conductive wire the resistance, $R$, is expected to change with stretch, $\lambda$, as ${R}/{R_{0}}=\lambda^2$, where $R_{0}$ is the resistance of the undeformed conductor. \cite{keplinger2013stretchable}
For a uniaxially stretched parallel-plate capacitor, the capacitance, $C$, is predicted to change as $C/C_{0}=\lambda$, where $C_{0}$ is the capacitance of the undeformed sensor. \cite{bartlett2016rapid}
Alternative stretching modes can also be analysed: for example, under equibiaxial stretching the capacitance is predicted to change as $C/C_{0}=\lambda^4$. \cite{sun2014ionic}
With these results, sensing can be achieved, even on delicate biological materials like a live rose petal (e.g. Figure \ref{fig:applications}H).

Detecting damage in soft robotics is also critical for long term durability and reliability.\cite{bartlett2019self}
Liquid-metal composites can detect the structural damage by by combining a sensing network on top of a sample, consisting of LM droplets embedded in elastomer.\cite{markvicka2019soft} 
When damage such as cutting, puncture, or impact is induced, LM droplets rupture and connect with the sensing network to localize and electrically report damage events.
By tuning the material composition, different damage modes can be detected and distinguished such as puncture or localized pressure.\cite{bartlett2019soft} 
Alternatively, optical detection of damage can also be achieved with liquid microcapsules embedded in silicones that change colour when damage is induced.\cite{li2016autonomous}

\subsubsection{Rigidity tuning}
Beyond passive changes to the elastic moduli of composites, functional liquids can also be used to produced solids with tunable stiffness.
The ability to dynamically change rigidity can enable morphing systems, human-machine interfaces, and vibration control.\cite{wang2018controllable} 
For example, rigidity tuning can be achieved through phase-change systems using materials that solidify near room temperature (e.g. paraffin or low melting point alloys like Field's metal).
These can be deployed as embedded droplets \cite{buckner2019enhanced}, in filled foam systems \cite{van2016morphing} as well as structured systems,\cite{deng2020multifunctional, Shan2013IJHMT, shan2013soft} to produce large changes in stiffness with temperature ((Figure \ref{fig:property}A)). 
This approach has also been used to produce stress-controlled stiffening by embedding under-cooled liquid droplets of Field's metal in a soft elastomer. In this case, solidification is triggered by mechanical deformation, increasing stiffness by 300\%. \cite{chang2018}
Fine-tunable control can also be gained by using embedded magnetorheological fluids (MRFs), where stiffness can be tuned with the application of different magnetic fields. \cite{testa2019}

\subsection{Soft electronics and wearables}
SLCs can be used to create soft electronic components such as highly stretchable wires, resistors, and capacitors (e.g. Figure \ref{fig:applications}G-K).\cite{Thrasher2019,Bartlett2016,li2019soft,mohammed2017all}
These systems are based on electrically conductive liquids such as EGaIn LM or ionic conductors such as ionic liquids,\cite{chen2018transparent,yang2015ionic} or water containing salt ions.\cite{yang2015ionic}
These soft electronic materials have shown exceptional combinations of soft mechanical response with diverse electrical functionality.
Fabrication routes are diverse, including various deposition techniques, soft lithography, microfluidics, and sintering of films of LM droplets by writing/scribing in the desired location.\cite{kramer2013masked,khoshmanesh2017liquid,martin2019heat,hirsch2016intrinsically,kim2017size}
In the latter case, the unique material architecture of conductive LM droplets also creates a self-healing response in the event of damage by rupturing adjacent droplets to form new connections and reroute the electrical signals (Figure \ref{fig:applications}D).\cite{Markvicka2018}

SLCs can also have excellent compatibility with human skin due to their softness and extensibility.
Thus, they have found use as wearable devices which can be used for applications including monitoring of various physiological signals, tracking motions, and harvesting energy (e.g. Figure \ref{fig:applications}L-Q). 
For example, Figure \ref{fig:applications}L shows a sweat-harvesting microfluidic patch that can analyzed to monitor hydration state or various chemical markers (Figure \ref{fig:applications}L). \cite{koh2016soft,choi2017thin}
Figures \ref{fig:applications}M,Q show SLC sensors that monitor hand position based on changes in resistance or capacitance. \cite{Tutika2019liquid, boley2015mechanically}
Furthermore, TENGs can be combined with LM composites to harvest energy for personal-health monitoring at low temperatures. \cite{malakooti2019liquid}
In some cases, traditional rigid electronics may still be required, and in this case, conductive liquids can be used as strain isolation layers, to decouple soft and stiff components to improve robustness in wearable devices.
This has been demonstrated using ionic liquids and soft silicones -- an appropriate pairing due to the negligible vapour pressure of ionic liquids and their extremely low solubility in silicone, which can effectively eliminate evaporation and leakage. \cite{ma2017soft}

\subsection{Chemical and biological applications}

There are is a great range of situations where SLCs are specifically designed - or make use of - chemical functionality (e.g. Figure \ref{fig:applications}R-U).
For example, as previously mentioned, encapsulation of reactive liquids is a very versatile technique for creating a self-healing composite.
As illustrated in Figure \ref{fig:applications}R,S when the composite is broken, the liquids release, and polymerise or otherwise initiate chemical reactions that fill in the developing crack.\edits{\cite{jones2014autonomic}}\cite{yin2007,white2001,xiao2009,hager2017}
This has been used demonstrated in various solids ranging from ceramics, to asphalt, concrete and a range of polymers. \cite{hager2017}

Biological cells also make use of liquid droplets, encapsulated in their cytoskeleton, to achieve a high degree of control over their internal chemical reactions. \cite{shin2017}
Cells form a variety of different types of $O(\mu\mathrm{m})$-sized intracellular droplets by liquid-liquid phase separation of proteins(Figure \ref{fig:applications}V).
Dissolved chemicals in the cell then either tend to concentrate in these droplets, or remain dissolved outside of them.
If all the necessary components of a chemical reaction are concentrated inside a droplet, the increase in local concentration can significantly increase reaction rates.
On the other hand, if a necessary component of a reaction is sequestered in a droplet, away from other components, then the reaction can be effectively stopped until something causes the droplet to dissolve, at which points reactants flood the cell.
Interestingly, recent work suggests that the elasticity of the surrounding network can play an important role in droplet behaviour. \cite{shin2018}

There are also many situations where it's important to sequester biological molecules in soft materials.
This could be for using the soft materials to encapsulate the biological molecules to studying degradation or reactions (Figure \ref{fig:applications}W).\cite{faccio2019}
It is also important for drug delivery.
About half of recently developed drugs are extremely hydrophobic. \cite{josef2010}
Thus they cannot simply be added into an aqueous system, as they will simply aggregate.
A novel approach to address this is to dissolve the drug in microscopic oil droplets in a hydrogel, and then use this as a hydrophilic delivery vehicle.
By controlling the hydrogel properties it is possible to tune factors such as drug release rate, and feel (for topical applications). \cite{josef2010,singh2014,wang2018cinnamon}

\section{Conclusions and future directions}

Although there has been seen significant progress in SLCs, we can see multiple fruitful directions for future work.
Although not an exhaustive list, we anticipate these three areas being of particular importance:

\textbf{Materials:}
From a Materials Science perspective, it is important to identify gaps and weaknesses in current material properties, and to fill these with novel composites.
We have discussed how SLCs already fill many gaps (for example in making stretchable conductors).
However, we are certain that SLC properties can be pushed even further than the current limits in Figure \ref{fig:megaproperties}, and that there are still novel properties to be uncovered.
\edits{This could come from exploring novel material combinations and geometries, from pushing these principles to smaller scales, or from combining multiple functionalities into materials.
Currently, scaleable techniques to organize liquid inclusions into predetermine arrangements are lacking and represent a rich future opportunity for controlling properties.
For example, optical properties depend strongly on inclusion organisation (e.g. amorphous vs periodic \cite{wiersma2013disordered}) so better control of this may lead to novel behavior.
One way to create nano-SLCs, is via the grafting of  polymer brushes on solid surfaces to effectively give a stable, liquid coating on the surface. These are well known to modify surface properties such as adhesion and wetting, and incorporating such nanoscopic liquid domains in composites (for instance as coatings on the surface of filler particles) offers many opportunities to create novel properties (e.g. \cite{hui2014surface}).
We also foresee that combining multiple different liquid phases could give multi-functional properties, for example coupling electrical or thermal activation with magnetic or optical stimuli.
This could even be combined with materials intelligence where the mechanical state or material configuration could program a functional, actuation, or computational response.
In addition, we note that very compelling evidence of the potential of SLCs, is the fact that Nature commonly uses such materials for a range of purposes, and can achieve (among other things) mechanical and optical properties and self-healing that out-perform most synthetic materials.}

\edits{Although we have discussed many useful properties and applications of SLCs, there are still many existing weaknesses that also need to be addressed.
Many SLCs utilize viscoelastic elastomers, such as silicones, that dissipate energy when they are rapidly stretched, meaning that they can not be used at too high speeds.
They can also suffer from low resistance to fatigue upon repeated cycling, or degradation upon exposure to harsh conditions such as UV light or salt-water environments.
Some SLCs need to be stabilised against evaporation of the liquid phase, or may not bond well to certain surfaces when they are integrated into mechanical systems.
For electrical applications, challenges include the fact that high voltages are typically required to achieve actuation, high switching energies, and there are common trade-offs in terms of actuation stroke and blocking force.
Addressing these challenges is a key step towards increasing the suitability of SLCs for commonplace engineering applications.}

\textbf{Manufacturing:}
A second challenge is in taking these materials from the lab, and applying them at scale.
Many of these materials have potential industrial uses, but only a small number of them make it from the lab to true large scale (although it is worth noticing that there are a growing number of startups in the area).
To overcome this gap, work needs to be done to take the most useful materials, and develop scaleable methods for their production.
This will likely require redesign of existing systems using practical materials -- for example in terms of their price, availability and biocompatibility, and sustainability.
\edits{Although a variety of novel fabrication techniques have been proposed, there are still many challenges remaining in the scaleability.  
Roll to roll fabrication processes and large batch processes currently used in industry to create printed materials and electronics and emulsion based material processing could be particularly valuable for future SLC materials.} 

\textbf{Theory:}
Finally, alongside the material development, there are a host of accompanying theoretical challenges.
We need theory to understand novel physical phenomena that arise in soft SLCs.
Theory can also help predict novel material combinations with useful properties, and indicate the potential achievable limits of materials.
It can investigate and understand novel, small-scale phenomena that arise as domain sizes shrink to the colloidal- or nano-scale.
Furthermore it can guide the design of materials, both in terms of microstructural geometry, and in terms of macroscopic shape, so as to optimise their use.
This is no mean task, given the huge parameter space. This also presents an opportunity for emerging approaches in artificial intelligence and machine learning to contribute to design. As first principal calculations can be challenging in multiscale, multimaterial systems such as SLCs, AI could provide a tool to optimize material design and fabrication.

\textbf{Closing remark:}
Despite having a long history, the concept of soft, solid-liquid composites is re-emerging as a material paradigm for the future.
The topic naturally brings together scientists from across fields including Chemistry, Soft Matter Physics, Soft Engineering, Biology, and Applied Mathematics, and we hope this review inspires more of them to contribute to the growing area.

\section*{Conflicts of interest}
There are no conflicts to declare.

\section*{Acknowledgements}
R.T. and M.D.B acknowledges support from a 3M Nontenured Faculty Award. J.Y.K. is supported by the MOTIE in Korea, under the Fostering Global Talents for Innovative Growth Program supervised by the KIAT (P0008746). R.W.S. is supported by the Swiss National Science foundation (grant 200021-172827).

\section{Appendix}
To calculate the effective stiffness of a composite of LM droplets embedded in a soft gel, we consider the system of an incompressible liquid droplet of radius $R$, embedded in an incompressible solid with Young's modulus $E_s$. There is a thin, interface layer (the metal oxide), with thickness $h\ll R$, Young's modulus $E_i$, and Poisson ratio $\nu_i$.
This system can be approximated by a droplet in a solid with surface elasticity at the interface, with surface elastic Lam\'{e} constants \cite{style2017}
\begin{equation}
    \lambda^s=\frac{E_i \nu_i h}{1-\nu_i^2},\quad \mu^s=\frac{E_i h}{2(1+\nu_i)}.
\end{equation}

Using the results of Duan et al., \cite{duan2007} we can then calculate the effective stiffness of an elastic particle that would have equivalent behaviour to the droplet and oxide layer as
\begin{equation}
    E_{\mathrm{eff}}=\frac{E_i h}{2R}\left(\frac{5 E_i h+2 E_s R (7-5\nu_i)}{E_i h (7+5\nu_i)+10 E_s R(1-\nu_i^2)}\right) \approx \frac{5 E_i h}{2 (7+5\nu_i) R},
\end{equation}
where the last approximation is the limit for $E_i h/E_s R \gg 1$.
Thus in this limit, the droplet effectively behaves as a particle with stiffness $\sim E_i h/R$.
Finally, we can use this to calculate the composite stiffness, by inserting it into equation (\ref{eqn:Ec}).

For gallium oxide, $E_i\approx 200$ GPa, $\nu_i\approx 0.3$, and $h\approx 1$ nm. \cite{farrell2018control}
Using these values, we see that droplets of radius $10\mu$m will stiffen solids with $E_s<6$ MPa.


\end{document}